\documentclass[aps,pre,reprint,superscriptaddress,footinbib,showpacs,floatfix]{revtex4-1}

\usepackage{comment}
\usepackage{graphicx}
\usepackage{latexsym}
\usepackage{xcolor}
\usepackage{amsmath}
\usepackage[caption=false]{subfig}
\usepackage{multirow}
\usepackage{booktabs}
\usepackage{textcomp}
\usepackage{placeins} % Makes \FloatBarrier available!
\usepackage{float}
\usepackage{mathtools}
\usepackage{array}
\usepackage{tabularx}
\usepackage{ulem} % Use \sout{} for strikethrough text.
\usepackage{amsfonts,amssymb}
\usepackage{accents}
\graphicspath{{images/}}

\newcommand{\ec}[1]{\textcolor{black}{#1}}

\newcommand{\be}{\begin{equation}}
\newcommand{\ee}{\end{equation}}
\newcommand{\bea}{\begin{eqnarray}}

\begin{document}

\title{Mechanisms of DNA-mediated allostery}

\author{Midas Segers}
\affiliation{Soft Matter and Biophysics, KU Leuven,
Celestijnenlaan 200D, 3001 Leuven, Belgium}
\author{Aderik Voorspoels}
\affiliation{Soft Matter and Biophysics, KU Leuven,
Celestijnenlaan 200D, 3001 Leuven, Belgium}
\author{Takahiro Sakaue}
\affiliation{Department of Physics and Mathematics, 
Aoyama Gakuin University, 5-10-1 Fuchinobe, Chuo-ku, 
Sagamihara, Kanagawa 252-5258, Japan}
\author{Enrico Carlon}
\affiliation{Soft Matter and Biophysics, KU Leuven,
Celestijnenlaan 200D, 3001 Leuven, Belgium}

\date{\today}

\begin{abstract}
Proteins often regulate their activities via allostery - or 
action at a distance - in which the binding of a ligand at one binding site 
influences the affinity for another ligand at a distal site. Although less 
studied than in proteins, allosteric effects have been observed in experiments 
with DNA as well. In these experiments two or more proteins bind at distinct 
DNA sites and interact indirectly with each other, via a mechanism mediated 
by the linker DNA molecule. We develop a mechanical model of DNA/protein 
interactions which predicts three distinct mechanisms of allostery. 
Two of these involve an enthalpy-mediated allostery, while a third 
mechanism is entropy driven. We analyze experiments of DNA allostery and 
highlight the distinctive signatures allowing one to identify which of the 
proposed mechanisms best fits the data.
\end{abstract}

\maketitle

\begin{figure}[t]
\includegraphics[width=0.45\textwidth,angle=0]{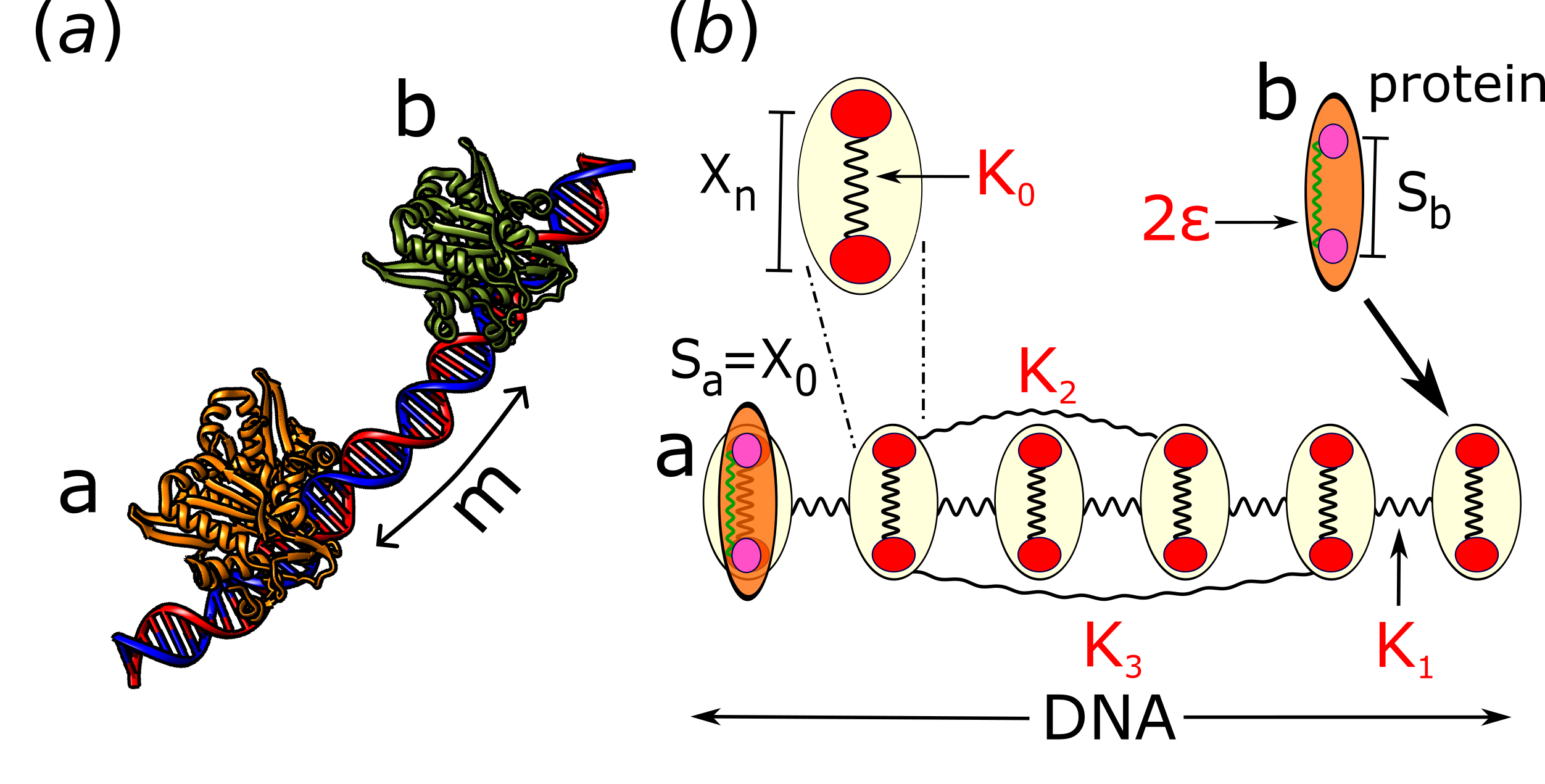}
\caption{(a) 
%  Allostery in DNA occurs when two proteins bound at 
%  distinct sites interact with each other via a DNA mediated interaction.
DNA-mediated interaction for two bound proteins separated by a linker
molecule of $m$ base pairs.
%  The length of the linker DNA, measured from the nearest edges 
%  of the two proteins, is $m$ base pairs.
(b) Model of DNA allostery. The DNA substrate (yellow) is described as
a set of variables $X_n \equiv u_n + \bar{l}$ 
defined at each base-pair site, 
with $\langle X_n \rangle = \bar{l}$
the equilibrium value. These variables are 
characterized by a local stiffness ($K_0$) and distal couplings 
($K_1$, $K_2$, \ldots), with energy given by \eqref{eq:model}. 
Upon binding, the protein, (orange blob) modifies the local mechanical 
properties of the DNA substrate. The distal couplings carry the signal 
to distinct sites. The schematic plot in (b) shows a protein 
interacting with a single DNA site. The more realistic case of proteins 
binding to several DNA sites is also considered.}
\label{fig:allo_model}
\end{figure}

The term ``allostery'' indicates an action at a distance in biological
macromolecules where the binding of a ligand at one site modifies the
binding of another ligand at a distinct site.  Many proteins regulate
their activities via allostery \cite{guna04}, through mechanisms that
are not fully understood and presently debated, see e.g.~\cite{woda19}.
Although hitherto most of the focus has been on proteins, allosteric
effects have been observed in DNA as well \cite{kim13,rose21} and
discussed in models and simulations \cite{drsa14,sing18,bala18}.
In DNA allostery two or more proteins binding at distinct sites
interact with each other through some signal carried by the linker
DNA, see Fig.~\ref{fig:allo_model}(a). Experiments show that the
interaction is weakly dependent on the DNA sequence \cite{kim13,rose21},
suggesting that allostery may be described by a homogenous DNA model. The
interaction is strongly attenuated if one of the two strands is cut (DNA
nicking), which shows that allostery requires an intact DNA molecule.
This and other experimental evidences \cite{kim13,rose21} show that the
interaction is transmitted through DNA, and not via direct (electrostatic)
or solvent-mediated effects. A model of force-induced allostery was
discussed in \cite{rudn99}. However, this mechanism does not apply to
experiments in which DNA is not under tension \cite{kim13,rose21}.

In all generality, the total free energy for the system consisting of
DNA and two bound proteins, separated by a linker sequence of $m$ base
pairs, is
\begin{equation}
F_{ab} = F_0 + \Delta F_a + \Delta F_b + \Delta\Delta F_{int}(m)
\label{eq:Fen_exp}
\end{equation}
where $F_0$ is the bulk contribution from DNA in absence of bound
proteins, $\Delta F_a$ and $\Delta F_b$ are the excess free energies
when only one of the two proteins, either ``a'' or ``b'', is bound. The
interaction term, $\Delta\Delta F_{int}(m)$, is the excess free
energy when both proteins are bound which vanishes as $m \to \infty$.
If $\Delta\Delta F_{int}<0$ the simultaneous binding of the two proteins
leads to a net decrease of the total free energy (cooperative binding).
If $\Delta\Delta F_{int}>0$, the simultaneous binding is destabilized.
%%%%%%%%%%%%%%%%%%%%%%%%%%%%%%%%%%%%%%%%%%%%%%%%%%%%%%%%%%%%%%%%%%%% 
%%%%%%%%%%%%%%%%%%%%%%%%%%%%%%%%%%%%%%%%%%%%%%%%%%%%%%%%%%%%%%%%%%%%  
We introduce a model which predicts three distinct mechanisms
of allostery, corresponding to different forms of $\Delta\Delta
F_{int}$. We introduce collective variables $X_n$ at each base pair
position $0 \leq n < N$, which are local reaction coordinates associated
to DNA-protein binding.  We define $\bar{l} \equiv \langle X_n \rangle$,
the equilibrium value, and $u_n \equiv X_n - \bar{l}$. At base pair
level, DNA deformations are described by several coarse-grained
coordinates like the $12$ canonical ones (twist, roll, tilt, rise,
\ldots) of the rigid base model \cite{olso01}. In our model $u_n$ could
be one of these coordinates, a combination thereof, or any other local
deformation parameter. Experimental data, discussed further, put %some
constraints on the properties of $u_n$ which gives insights on candidate
allostery-carrying variables.
In the $u_n$ the energy of free DNA %molecule is given by
is
\begin{equation}
H_0 = \frac{1}{2} \sum_{n=0}^{N-1} 
\left[ K_0 u_n^2 + \sum_{p=1}^L K_p (u_n u_{n+p}+u_n u_{n-p}) \right]
\label{eq:model}
\end{equation}
%  We consider $u_n$ as quadratic variables 
which is quadratic with local stiffness $K_0$
and distal couplings $K_p u_n u_{n+p}$ (Fig.~\ref{fig:allo_model}(b))
here assumed to extend to a finite range $L$.
Distal couplings naturally arise  from the collective nature of $u_n$ and are 
indeed observed in simulations 
\cite{lank00,noy12,esla11,skor21,fosa23}. They typically extend to a few
flanking nucleotides (truncated at a distance $L$ in \eqref{eq:model}) 
and the strength and decay of the interactions depend on 
the coarse-grained variable considered \cite{skor21}. %Here we assume a finite
% coupling distance $L$, which can be taken arbitrary large in the model. 
Distal
% , as shown below. 
% Such 
couplings are essential 
to generate %allosteric interactions. 
allostery.
The model \eqref{eq:model} is coarse-grained 
with one degree of freedom per DNA site %, but it has the advantage that it can be 
and it can be solved analytically. 
% Translation invariance implies that \eqref{eq:model} can be written as
Using periodic boundary conditions ($u_N \equiv u_0$) we write \eqref{eq:model} as
\begin{eqnarray}
H_0 &=& \frac{1}{2N} \sum_q \widetilde{K}_q \left| {\cal U}_q \right|^2
\label{eq:model2}
\end{eqnarray}
where we introduced the discrete Fourier transforms
\begin{equation}
{\cal U}_q = \sum_{n=0}^{N-1} e^{-2\pi i n q/N} u_n, \qquad
\widetilde{K}_q = 
\sum_{n=-L}^L  K_n \cos \left( \frac{2 \pi n q}{N} \right)
\label{def:FT}
\end{equation}
with $q$ an integer, $K_{-p}=K_p$ and $-N/2 < q < N/2$
\footnote{Models as \eqref{eq:model} were used in the literature
to describe large scale packing of chromosomal DNA, see A. Amitai
and D. Holcman, Phys. Rev. E 88, 052604 (2013) and K. Polovnikov,
S. Nechaev, and M. Tamm, Soft Matter 14, 6561 (2018). Some of these
models considered a slowly algebraic decay of the off-site couplings
$K_p \sim p^{-a}$ (and $L \to \infty$), which gives a Fourier space
stiffness ${\protect{\widetilde{K}_q}}$ non-analytic in $q=0$. We consider here
couplings extending to a finite range $L$, hence such non-analytic
behavior will not occur.}. In absence of distal couplings ($K_p=0$ for
$p \geq 1$) the $q$-stiffness $\widetilde{K}_q=K_0$ is constant, thus a
$q$-dependence of $\widetilde{K}_q$ reflects the existence of couplings
between distal sites.  Note that the couplings $K_p$ can take any value
as long as $\widetilde{K}_q >0$ for real $q$ (stability condition).

We consider first proteins interacting with a single DNA site.  An unbound
protein is thus described by a single collective variable $S$, with
average $\langle S \rangle = \bar{s}$ and energy $H_p = \varepsilon \left(
S - \bar{s} \right)^2$. The binding to DNA (at site $n=0$) forces the
corresponding collective variables to assume the same value $S=X_0=
\bar{l}+u_0$ so that the total energy of DNA and protein ($H_0+H_p$)
takes the form
\begin{equation}
H = H_0 + \varepsilon \left[ u_0^2 + 2(\bar{l}-\bar{s}) u_0 + (\bar{l}-\bar{s})^2\right],
    \label{bind1}
\end{equation} 
omitting a constant binding energy which does not influence $\Delta\Delta
F_{int}$.  Equation \eqref{bind1} shows that protein binding introduces
perturbation ``fields'' proportional to $u_0$ and $u_0^2$.  If the
equilibrium value of the collective coordinates of DNA and protein
coincide ($\bar{s} =\bar{l}$), the linear term vanishes and only the
term proportional to $u_0^2$ ``survives'' in \eqref{bind1}.  Conversely,
if $|\bar{l}-\bar{s}|$ is large one can neglect the quadratic term
contribution \footnote{ The linear term is dominant when $|l_0-s_0|^2 \gg
\langle u_0^2 \rangle_0$}.  In the following we compute $\Delta\Delta
F_{int}$ for three different cases where the proteins induces a linear
or a quadratic field.

A quantity of central interest is the propagator 
\begin{eqnarray}
    S_m \equiv \frac{1}{N} \sum_q \frac{e^{2\pi i mq/N}}{\widetilde{K}_q}
    = \beta \langle u_0 u_m \rangle_{\! _0},
\label{defSm}
\end{eqnarray}
where $\beta =1/k_BT$ 
is the inverse temperature and $\langle .\rangle_{\! _0}$ indicates a thermal 
average with respect of $H_0$. Equation~\eqref{defSm} follows from the 
equipartition theorem 
\begin{eqnarray}
    \beta \langle {\cal U}_q {\cal U}_p \rangle_{\! _0} &=& 
    N \widetilde{K}_q^{-1}  \delta_{q,-p}.
    \label{eq:eqp}
\end{eqnarray}
Transforming the sum in \eqref{defSm} into an integral ($N \to \infty$ limit), 
one obtains the asymptotic behavior of $S_m$ from the leading pole, i.e. the 
solution of $\widetilde{K}_q=0$ with the smallest imaginary part. This equation 
cannot have solutions 
for real $q$ as stability requires $\widetilde{K}_q>0$.
In the most general case the leading pole has real and imaginary parts. 
Since $\widetilde{K}_q$ is real for real $q$ and symmetric in $\pm q$ (see \eqref{def:FT}) 
there are at least four poles, one of which is
\begin{eqnarray}
    \frac{2 \pi q_E}{N} \equiv \phi + \frac{i}{\xi_E}
\label{def:qE}
\end{eqnarray}
and the others are $-q_E$, $q_E^*$ and $-q_E^*$, where $ ^*$ denotes complex conjugation. 
The asymptotic behavior is governed by $q_E$
\begin{eqnarray}
    S_m &\stackrel{m \gg 1}{\sim}& \Gamma \cos(m \phi + \phi_0)  \, e^{-m/\xi_E}
    \label{Smgg1}
\end{eqnarray}
with $\phi_0$ a phase shift and $\Gamma$ a scale factor
\footnote{See Supplemental material at http:, which contains references
\cite{lank09,voor23,petk14,gromacs,pere07,ivan16,li19,jorg83,buss07,parr81,hess97,lieb23},
for details of mathematical derivations, simulations and additional
information about the models used.}.

\begin{figure}[t]
\includegraphics[width=0.48\textwidth,angle=0]{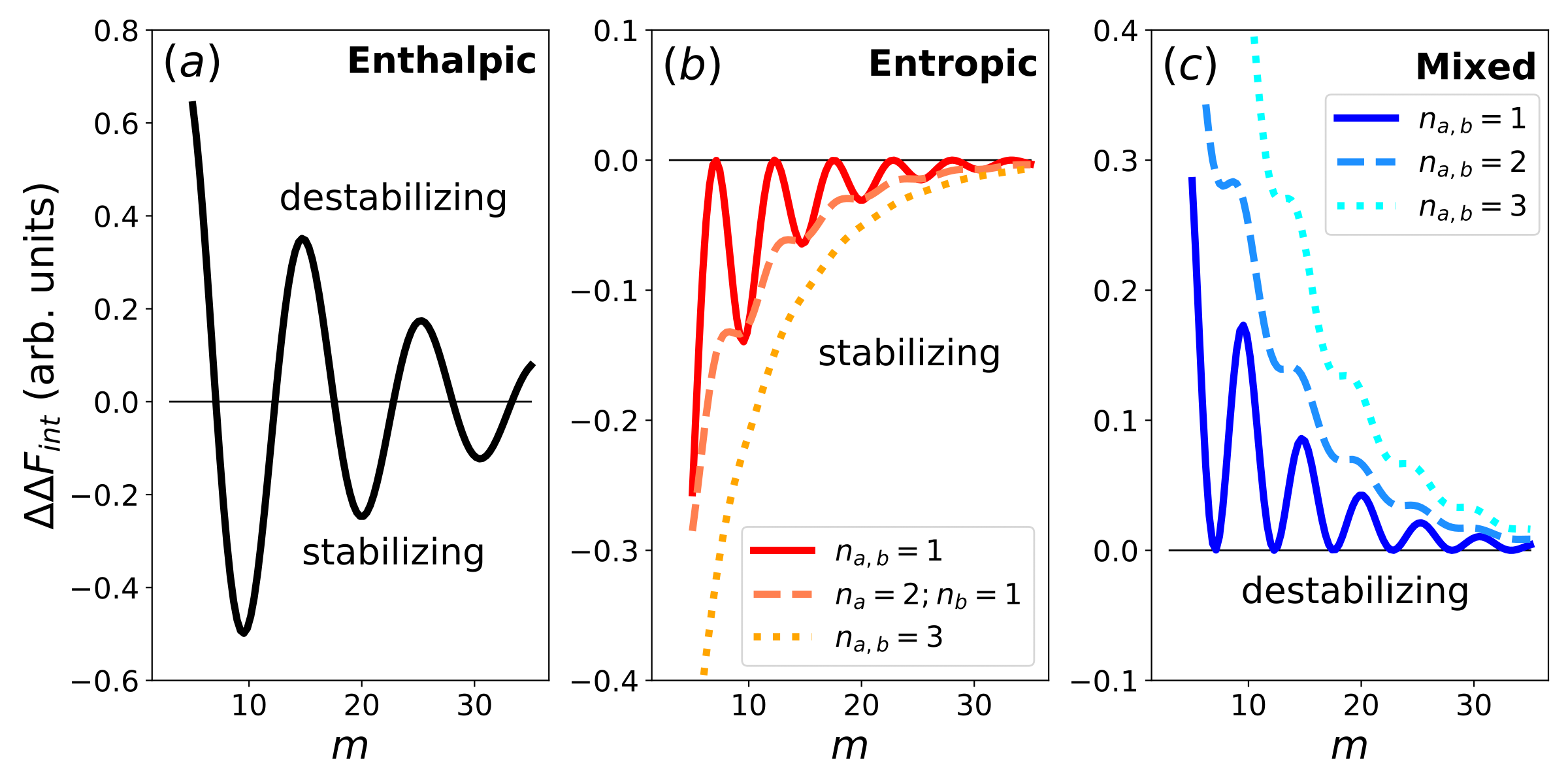}
\caption{Plots of $\Delta\Delta F_{int}$ vs. linker DNA length $m$ for
the three different mechanisms of DNA-mediated allostery proposed:
(a) enthalpic, (b) entropic and (c) mixed. An angular frequency
$\phi=2\pi/10.5$, corresponding to the periodicity of the DNA double
helix and $\xi_E=15$~bp were used. These parameters match those observed
in experiments (see Fig.~\eqref{fig:experiments}).  In the case (a)
the interaction is stabilizing/destabilizing depending on the values
of $m$ and the asymptotic oscillating behavior is universal, i.e. also
valid for interactions involving $n_a$, $n_b$ sites for protein a and
b, respectively.  In the cases (b) and (c) the interaction is always
stabilizing and destabilizing, respectively. In (b) and (c) the asymptotic
decay is non-universal, being dependent on the number of interacting
sites $n_{a,b}$ per protein.}
\label{fig:interaction}
\end{figure}

{\sl Enthalpic allostery --} 
We consider first
\begin{equation}
H^E = H_0 - h (u_0 + u_m) 
\label{pert1}
\end{equation}
with $h = -2 \varepsilon (\bar{l}-\bar{s})$, following \eqref{bind1}.
We find 
\cite{Note3}
\begin{eqnarray}
    \Delta\Delta F_{int}^E &=&  -h^2 \, S_m
    \label{DDF_enthalpy}
\end{eqnarray}
which, being temperature independent, describes an interaction of
enthalpic origin \footnote{The coarse-grained variables $u_n$ derive from
an underlying partial integration of  microscopic degrees of freedom.
This may lead to a temperature dependence of the couplings $K_n$.
Here enthalpic/entropic interactions refer to this coarse-grained level
of description. Our focus is on the sign of $\Delta\Delta F_{int}$
rather than its temperature dependence.}. The fields in $0$ and
$m$ shift the equilibrium values of $u_0$ and $u_m$ and the distal
couplings $K_1$, $K_2$, \ldots propagate this perturbation to flanking
sites, leading to $\langle u_n \rangle \neq 0$. Asymptotically the
interaction decays via damped oscillations, see \eqref{Smgg1}. We refer
to $\xi_E$ in \eqref{Smgg1} as the enthalpic allosteric length.  The
interaction stabilizes or destabilizes the simultaneous protein binding
depending on their distance $m$, see Fig.~\ref{fig:interaction}(a).
The calculation can be generalized to protein-DNA couplings involving
more than one site, i.e. of the type $\sum_{k=0}^{n_a-1} h_k u_k +
\sum^{n_a+n_b-2}_{l={n_a-1}} h_{m+l} u_{m+l}$ where $n_a$ ($n_b$) are
the number of sites to which first (second) protein binds and $m$ is the
number of base pairs separating the nearest edges of the two proteins.
The asymptotic decay remains of the form \eqref{Smgg1} which is universal
\cite{Note3}.
%%%%%%%%%%%%%%%%%%%%%%%%%%%%%%%%%%%%%%%%%%%%%%%%%%%%%%%%%%%%%%%%%%%%  
%%%%%%%%%%%%%%%%%%%%%%%%%%%%%%%%%%%%%%%%%%%%%%%%%%%%%%%%%%%%%%%%%%%%  

{\sl Entropic allostery -} We consider next
\begin{equation}
H^S = H_0 + \varepsilon \left( u_0^2 + u_m^2 \right)
\label{pert2}
\end{equation}
Differently from the enthalpic case, here $\langle u_n \rangle=0$
for all sites. We find \cite{Note3}
\begin{equation} 
    \Delta\Delta F_{int}^ S = \frac{k_BT}{2}
    \log \left[ 1- \left(\frac{2\varepsilon S_m}{1+2\varepsilon S_0} \right)^2
    \right]
    \label{DDF_entropy}
\end{equation}
The interaction is of entropic origin $\Delta\Delta F_{int}^S = 
- T \Delta\Delta S \leq0$,
implying a net increase in entropy 
when both proteins are bound (cooperative binding). 
This can be understood as follows. 
The local stiffening to $K_0+2\varepsilon$ at sites $0$ and $m$ induces
an entropy reduction in two regions surrounding the two perturbed
sites. When $m$ is sufficiently small, the two regions overlap which
leads to a net entropy gain, hence $\Delta\Delta S > 0$. This is
reminiscent of entropic attractions observed in soft condensed matter
systems, such as polymer-colloid mixtures \cite{witt10}.  In the limit
$m \gg 1$, $\Delta\Delta F_{int}^S$ vanishes as $S_m^2$. This implies
(Eq.~\eqref{Smgg1}) a decay length which is half of the enthalpic
allosteric length $\xi_S = \frac{1}{2} \xi_E$, and an oscillating
prefactor proportional to $\cos^2 (m \phi + \phi_0)$, as shown in
Fig.~\ref{fig:interaction}(b), red solid line.  We extended the analysis
of $\Delta\Delta F_{int}^S$ for protein-DNA contacts at more than one
site. We consider first $\varepsilon (u_0^2 + u_1^2 + u_{m+1}^2)$, which
can be solved analytically (\cite{Note3}, Eq.~(S43)) shown as dashed
line in Fig.~\ref{fig:interaction}(b). Unlike \eqref{DDF_entropy},
this extended binding case contains terms proportional to $S_m^2$,
$S_{m+1}^2$ and $S_m S_{m+1}$, each oscillating but with different
phases. Figure~\ref{fig:interaction}(b) (dotted) shows $\Delta\Delta
F_{int}^S$ for an interaction term $\varepsilon (\sum_{l=0}^2 u_l^2 +
\sum_{k=m}^{m+2} u_k^2)$ in which each protein couples to a block of
$3u$'s. There is in this case a very weak modulation of the exponential
decay. Summarizing, the asymptotic behavior for generic quadratic
interactions is
\begin{eqnarray}
    \Delta\Delta F_{int}^S \sim f(m) \,e^{-2m/\xi_E}
\label{eq:non-univ}
\end{eqnarray}
with a non-universal prefactor $f(m) \leq 0$, which depends on details
of the protein-DNA bindings (unlike the universal behavior of the
enthalpic case).

{\sl Mixed allostery -} Finally, we consider the mixed case
\begin{eqnarray}
    H^M = H_0 - h u_0 + \varepsilon u_m^2
    \label{def:mixed}
\end{eqnarray}
for which we find \cite{Note3}
\begin{equation}
     \Delta\Delta F_{int}^M = 
     \frac{\varepsilon h^2 S_m^2}{1+2\varepsilon S_0}
     \label{DDF_mixed}
\end{equation}
which is positive and thus a destabilizing interaction term.  It is
temperature independent and thus of enthalpic nature. The term $-h
u_0$ produces a $\langle u_n \rangle \neq 0$, which contributes to the
enthalpic part, but we find no entropy change in this model.
As $\Delta\Delta F_{int}^M$ depends on $S_m^2$ the asymptotic is
very similar to the entropic case, with decay length $\xi_M = \frac{1}{2}
\xi_E$ and oscillations proportional to $\cos^2(m\phi+\phi_0)$. As in
the entropic case, interactions to more than one site
lead to a decay of the type \eqref{eq:non-univ}, with $f(m) \geq 0$.
We note that \eqref{DDF_enthalpy} and \eqref{DDF_entropy} (but not
\eqref{DDF_mixed}) were also derived in a study of interactions of point
defects in fluctuating membranes \cite{netz97}. Their applicability is
general and not limited to a one dimensional chain.

%%%%%%%%%%%%%%%%%%%%%%%%%%%%%%%%%%%%%%%%%%%%%%%%%%%%%%%%%%%%%%%%%%%%  
\begin{figure}[t]
\includegraphics[width=0.48\textwidth,angle=0]{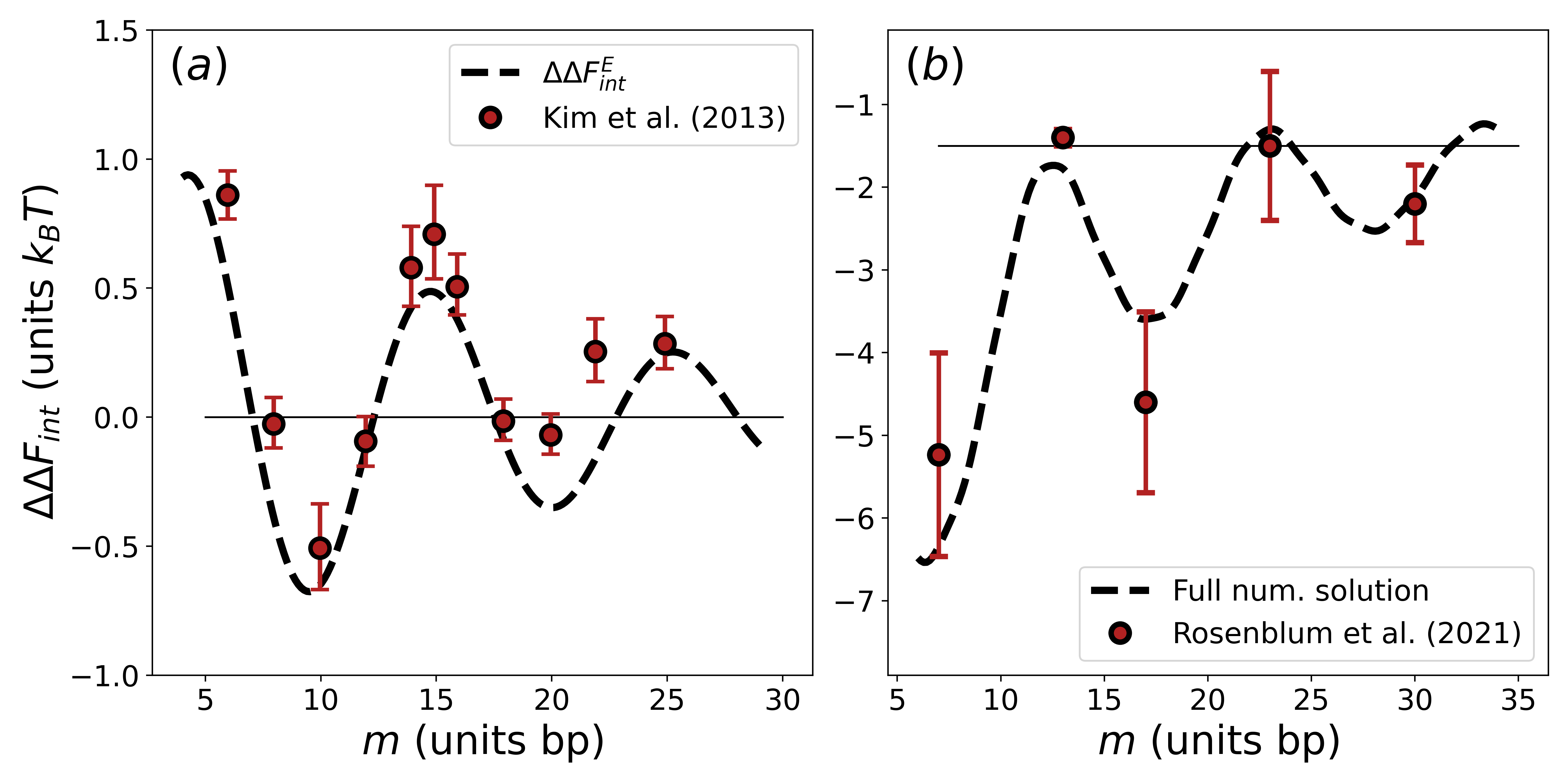}
\caption{(a) Symbols: Experimental data of interaction free energies for
(BamHI-GRDBD) \cite{kim13}. Oscillating sign in $\Delta\Delta F_{int}$
indicates an enthalpic type of allostery. Dashed line: Fit to the
asymtptotic expression \eqref{Smgg1} ($\phi=2\pi/10.5$, $\xi_E=15$bp).
(b) Symbols: Experimental data for $\Delta\Delta F_{int}$ for the
ComK system \cite{rose21}. The negative sign indicates an allosteric
interaction which is of predominant entropic. Dashed line: Full numerical
solution of $\Delta\Delta F_{int}(m)$ ($\phi=2\pi/10.5$, $\xi_E=22$bp)
for a global allostery model with extended perturbations \cite{Note3}.
As $\Delta\Delta F_{int}$ does not seem to converge to zero for large $m$,
we have addeed an asymptotic non-zero offset (dotted line).}
\label{fig:experiments}
\end{figure}
%%%%%%%%%%%%%%%%%%%%%%%%%%%%%%%%%%%%%%%%%%%%%%%%%%%%%%%%%%%%%%%%%%%%  

{\sl Experiments} -
In principle, one could distinguish the three scenarios in experiments
from the sign of $\Delta\Delta F_{int}$ (Fig.~\ref{fig:interaction}). Kim
{\sl et al.} analyzed the binding of several different pairs of
proteins on DNA \cite{kim13}. The binding free energy showed a decaying
oscillating behavior with alternating sign which is consistent with
an enthalpic allostery \eqref{pert1}, in agreement with the analysis
performed by other authors \cite{kim13,drsa14,drsa16,sing18}.  A fit to
\eqref{pert1} with the asymtpotic expression for $S_m$ \eqref{Smgg1}
is shown in Fig.~\ref{fig:experiments}(a).  A different system was
analyzed by Rosenblum {\sl et al.} \cite{rose21} who found DNA-mediated
allostery in the binding of bacterial transcription factors ComK. The
experiments showed a cooperative binding ($\Delta\Delta F_{int}<0$) for
varying spacer lengths, see Fig.~\ref{fig:experiments}(b). The negative
$\Delta\Delta F_{int}$ indicates an allostery of (predominantly)
entropic type.  The data are fitted (dashed line) against a model
containing both linear and quadratic terms, using a Monte Carlo fitting
procedure \cite{Note3}.  These fields act on several sites reflecting the
extended contact regions of the ComK-DNA interaction. The oscillating
component is due to the enthalpic part, which, as seen above, has a
universal oscillatory decaying behavior. In the fit the same value of
$\phi=2\pi/10.5$ as Fig.~\ref{fig:experiments}(a) was used.  Instead for
the correlation length we used $\xi_E=22$~bp, larger than the value
used in (a), possibly indicating that allosteric coupling is carried
over by different collective variables in the two cases.  We note that
the ComK have much larger $|\Delta\Delta F_{int}|$ than the data shown
in Fig.~\ref{fig:experiments}(a). It is possible that to quantitatively
describe the ComK data one would need to go beyond linear elasticity
(anharmonic effects) or to more complex multimodal models \cite{lope23}.
However, the harmonic model with linear and quadratic terms spanning
several sites generates a $\Delta\Delta F_{int}$ consistent with
experimental data.

%%%%%%%%%%%%%%%%%%%%%%%%%%%%%%%%%%%%%%%%%%%%%%%%%%%%%%%%%%%%%%%%%%%%  
\begin{figure}[t]
\includegraphics[width=0.48\textwidth,angle=0]{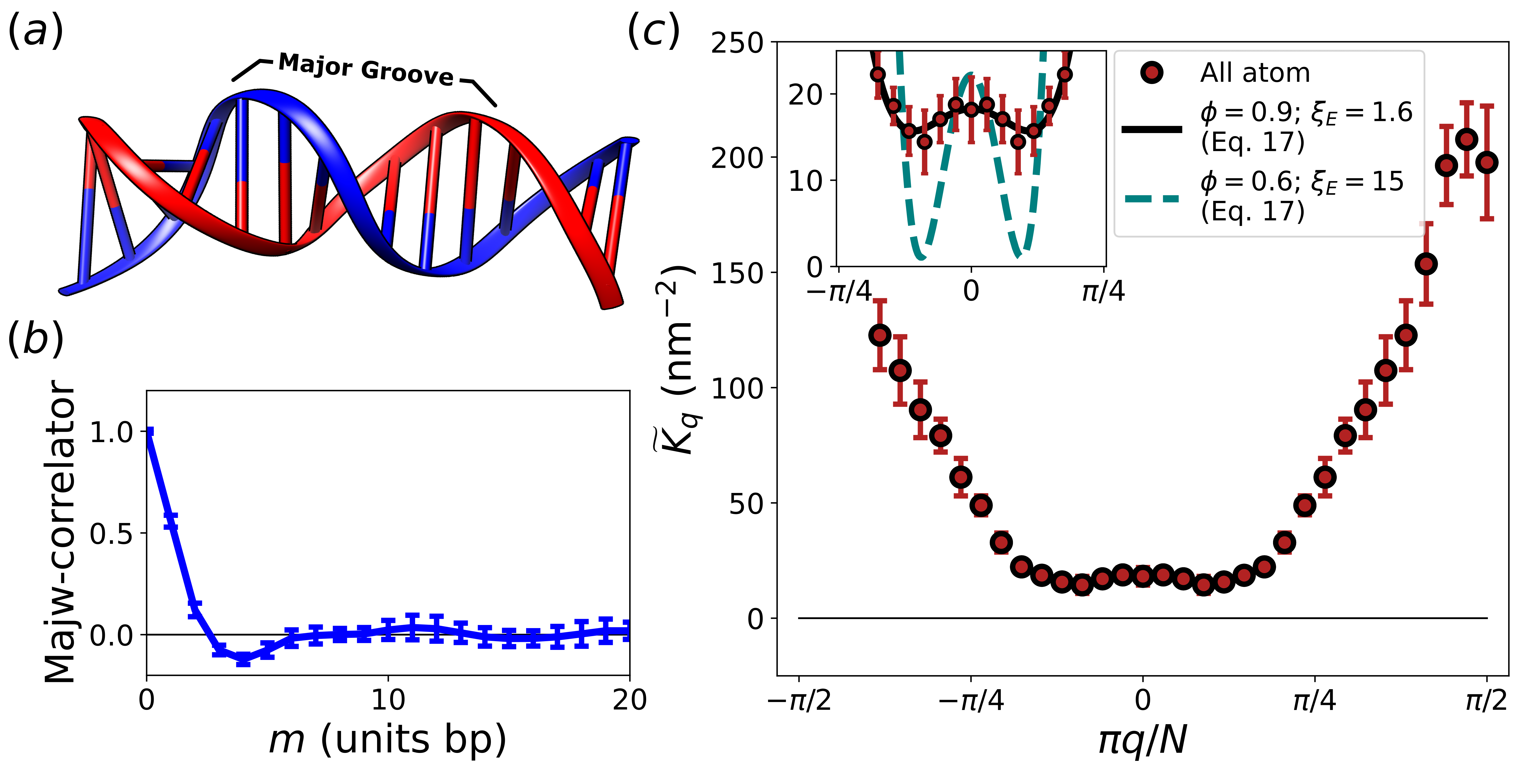}
\caption{(a) Schematic representation of the major groove width of
DNA, in practice the Curves+ \cite{lave09} definition is used.  (b)
Correlation function of the major groove width as obtained from all-atom
simulations obtained using the Curves+ software \cite{lave09}. (c)
Plot of the momentum-space stiffness for the major groove width obtained
from all-atom data. The inset coplots $\widetilde{K}_q$ for the minimal
model (Eq.~\eqref{def:Kq-minimal}) for two sets of parameters. The solid
line is a direct fit of the stiffness data. The dashed line uses $\phi$
and $\xi_E$ from a fit of the experimental $\Delta\Delta F_{int}$ data.
The error-bars in (b) and (c) indicate the standard deviation calculated
over 21 time windows of $10$~ns.}
\label{fig:simulations}
\end{figure}
%%%%%%%%%%%%%%%%%%%%%%%%%%%%%%%%%%%%%%%%%%%%%%%%%%%%%%%%%%%%%%%%%%%%  

{\sl Simulations} -
So far we have assumed a model with distal couplings generating
allosteric interactions, using a generic $u_n$.  In principle $u_n$
could be one of the several DNA local deformation modes used in the
rigid base model \cite{olso01}, or a combination thereof. However,
experiments indicate that allosteric interactions decay via damped
oscillations, which puts some constraints on $u_n$, as several variables
do not have this property. For instance, we can exclude pure bending or
twist modes as candidates for $u_n$, as the stiffness $\widetilde{K}_q$
for these coordinates does not produce oscillations with the desired
periodicity \cite{Note3}.  Prior work suggested the DNA major groove
width as mediating allosteric interactions \cite{kim13}.  However,
extensive $1$~$\mu$s all-atom simulations of a 33-bp sequence showed no
signature of periodicity in the groove width correlations \cite{drsa16}.
We have performed simulations using two different $44$-bp
sequences for $100$~ns, with major groove width calculated from
the algorithm Curves+ \cite{lave09} using the setup discussed in
\cite{skor21}. Our results for the normalized propagator $S_m/S_0 =
\langle u_0 u_m\rangle_{\! _0}/\langle u_0^2 \rangle_{\! _0}$ (with
$u_n$ the deviation of the major groove width from the equilibrium
value) are given in Fig.~\ref{fig:simulations}(b) and are in close
agreement with those reported in \cite{drsa16}.  To extract parameters
from simulations we have calculated the $q$-stiffness $\widetilde{K}_q$
from the equipartition relation \eqref{eq:eqp}, as recently done for
twist and bending deformations \cite{skor21,sege22}. A minimal model
with just four poles $\pm q_E$, $\pm q_E^*$ \eqref{def:qE} gives
\begin{equation}
\widetilde{K}_q = A (q^2-q_E^2)(q^2-q_E^{*2})= A(q^4 - \mu q^2 + \lambda^2)
\label{def:Kq-minimal}
\end{equation}
with $A$ a scale factor, $\mu=q_E^2+q_E^{*2}$ and $\lambda=|q_E|^2$.
We note that \eqref{def:Kq-minimal} is not of the form \eqref{def:FT},
but should be interpreted as the continuum long wavelength limit ($q\to
0$) of model \eqref{eq:model}.  Figure~\ref{fig:simulations}(c) shows
$\widetilde{K}_q$ as obtained from simulation data averaged over two
sequences (red circles). The double-well shaped curve is fitted, for small
$q$, to \eqref{def:Kq-minimal} giving $\phi=0.9$ and $\xi_E=1.6$ (solid
line in the inset of Fig.~\ref{fig:simulations}).  The former parameter
is close to the double helix periodicity $\phi=2\pi/10.5 \approx 0.6$,
while the latter appears to be quite small as compared to experimental
data which predict $\xi_E \approx 15$~bp. We show on the same inset a
plot of Eq.~\eqref{def:Kq-minimal} with the latter values for $\phi$
and $\xi_E$. We conclude that simulations of the major-groove width
qualitatively support the distal couplings model \eqref{eq:model}, but a
quantitative matching remains an open challenge, as pointed out earlier
\cite{drsa16,bala18}.  See \cite{Note3} for an extended discussion on
possible origins of these discrepancies.

{\sl Conclusions} -
We have studied a coarse-grained model which predicts three types of
allosteric DNA-mediated interactions.  One could distinguish between the
three cases  (or about the dominance of one of these) from the sign of
the interaction free energy and on its dependence on the length of the
DNA linker sequence separating the two protein-binding sites.  Prior
work \cite{kim13,drsa14,sing18} pointed to some examples of enthalpic
DNA-mediated allostery characterized by a free energy of oscillating
sign. We have argued here that recent ComK data \cite{rose21} show an
allostery which is of predominant entropic nature, as $\Delta\Delta
F_{int}<0$.  Entropic allostery (often referred as dynamic allostery)
was discussed in the case of proteins \cite{hawk04,dokh16}, but it
should manifest itself in DNA as well.  The model introduced here
predicts additionally a ``mixed'' allostery, obtained when coupling two
different proteins in which one exerts a linear field and the other a
quadratic one. This mixed allostery is of enthalpic nature and we are
not aware that such interaction was discussed in the protein literature.
Differently from the protein case, in DNA-mediated allostery one can vary
the spacer sequence length, probing the decay of $\Delta\Delta F_{int}$,
therefore the type of allostery (enthalpic, entropic or mixed) should
be easier to identify.  By varying the binding sites sequences one can
bring in close vicinity proteins of different types and which couple
differently to the DNA (e.g. predominantly via linear or quadratic
fields), thereby probing the three scenarios predicted by the model.

MS acknowledges financial support from Fonds Wetenschappelijk Onderzoek
(FWO 11O2323N). EC thanks the Aoyama-Gakuin University (Japan) and
Laboratoire C. Coulomb of the University of Montpellier (France), where
part of this work was done, for the kind hospitality. He acknowledges
financial support from the Japan Society for the Promotion of Science
(FY2020 JSPS).  Discussions with G. Bussi, H. Hofmann and G. Rosenblum
are gratefully acknowledged.

% \vspace{-3mm}
\bibliography{references}
%  \end{document}

%merlin.mbs apsrev4-1.bst 2010-07-25 4.21a (PWD, AO, DPC) hacked
%Control: key (0)
%Control: author (8) initials jnrlst
%Control: editor formatted (1) identically to author
%Control: production of article title (-1) disabled
%Control: page (0) single
%Control: year (1) truncated
%Control: production of eprint (0) enabled
%

\vfill\eject

\section*{Supplemental Information M. Segers et al. "Mechanisms of
DNA-mediated allostery"}

This document is organized as follows.  Section~\ref{sec:details_calc}
shows the details of the calculations of the exact expressions of
the interaction free energy $\Delta\Delta F_{int}$ for enthalpic
(\ref{subsec:enthalpy}), entropic (\ref{subsec:entropy}) and mixed
(\ref{subsec:mixed}) allostery. These are given in the framed equations
\eqref{SI:DDFenth2}, \eqref{SI:DDF} and \eqref{SI:DDF_mixed} and are
expressed as functions of the propagator $S_m$. These results are derived
for pointwise interactions at two single sites and are generalized to the
case of proteins interacting to multiple sites in~\ref{subsec:extended}.
Section~\ref{sec:propagator} discusses the asymptotic behavior of the
propagator, the allosteric length and its connection with DNA persistence
lengths.  Section~\ref{supp:fitComK} gives some details of the procedure
followed to fit the ComK data. Section~\ref{supp:AA} discusses details
of all-atom simulations.

\section{Enthalpic, Entropic and Mixed Allostery}
\label{sec:details_calc}

%%%%%%%%%%%%%%%%%%%%%%%%%%%%%%%%%%%%%%%%%%%%%%%%%%%%%%%%%%%%%%%%%%%%%%%%%%
%%%%%%%%%%%%%%%%%%%%%%%%%%%%%%%%%%%%%%%%%%%%%%%%%%%%%%%%%%%%%%%%%%%%%%%%%%
\subsection{Correlation functions in free DNA}
\label{subsec:correl}
%%%%%%%%%%%%%%%%%%%%%%%%%%%%%%%%%%%%%%%%%%%%%%%%%%%%%%%%%%%%%%%%%%%%%%%%%%
%%%%%%%%%%%%%%%%%%%%%%%%%%%%%%%%%%%%%%%%%%%%%%%%%%%%%%%%%%%%%%%%%%%%%%%%%%

Let us consider the free DNA model as given by %\eqref{eq:model} 
(2)
in the main text.  The equilibrium probability distribution $p(u_0,u_1
\ldots u_{N-1})$, due to the quadratic nature of the interactions, is a
multivariate gaussian distribution.  Integrating all variables but $u_0$
and $u_m$ gives a joint probability distribution for $u_0$ and $u_m$
which is still gaussian
\begin{equation}
p(u_0,u_m) \sim e^{-A(u_0^2+u_m^2)- 2 B u_0 u_m}
\label{SI:pu0um}
\end{equation}
with $A$ and $B$ some coefficients.
From integration of \eqref{SI:pu0um} we find
\begin{eqnarray}
    \langle u_0 u_m \rangle_{\! _0} &=& - \frac{1}{2} \, 
    \frac{ B}{A^2-B^2} 
    \label{SI:u0um}\\
    \langle u_0^2 \rangle_{\! _0} &=& \frac{1}{2} \, 
    \frac{A}{A^2-B^2}
    \label{SI:u02}
\end{eqnarray}
where we used the subscript ``0'' to stress that these are 
correlations of free DNA in absence of perturbing fields.
Using the inverse Fourier transform
\begin{equation}
u_n = \frac{1}{N} \sum_{q=0}^{N-1} e^{2\pi i qn/N} \, {\cal U}_q,
\label{def:un}
\end{equation}
we can compute the correlations directly
\begin{eqnarray}
    \langle u_0 u_m \rangle_{\! _0} &=& 
	\frac{1}{N^2} \sum_{q,q'} e^{2\pi i m q/N} 
    \langle {\cal U}_q {\cal U}_{q'}\rangle_{\! _0} \nonumber \\
    &=& \frac{k_BT}{N} \sum_q \frac{e^{2\pi i m q/N}}{\widetilde{K}_q}
    = k_BT S_m
    \label{SI:u0um-dir}
\end{eqnarray}
where the propagator $S_m$ follows the same definition of the main
text %\eqref{defSm}. 
(6).
To obtain \eqref{SI:u0um-dir} we have used the
equipartition relation
\begin{eqnarray}
    \langle {\cal U}_q {\cal U}_{q'} \rangle_0 &=& 
    N k_B T \widetilde{K}_q^{-1} \, \delta_{q,-q'}
\end{eqnarray}
Setting $m=0$ in \eqref{SI:u0um-dir} we get
\begin{eqnarray}
    \langle u_0^2 \rangle_{\! _0} &=&  
    \frac{k_BT}{N} \sum_q \frac{1}{\widetilde{K}_q} = k_BT S_0
    \label{SI:u02-dir}
\end{eqnarray}
Comparing \eqref{SI:u0um}, \eqref{SI:u02}  with \eqref{SI:u0um-dir}, 
\eqref{SI:u02-dir} we obtain the following expressions for $A$ and $B$
\begin{eqnarray}
    A &=& \frac{\beta}{2} \, 
    \frac{S_0}{S_0^2 - S_m^2} 
    \label{SI:defA}
\end{eqnarray}
\begin{eqnarray}
    B &=& \frac{\beta}{2} \, 
    \frac{-S_m}{S_0^2 - S_m^2}
    \label{SI:defB}
\end{eqnarray}

%%%%%%%%%%%%%%%%%%%%%%%%%%%%%%%%%%%%%%%%%%%%%%%%%%%%%%%%%%%%%%%%%%%%%%%%%%
%%%%%%%%%%%%%%%%%%%%%%%%%%%%%%%%%%%%%%%%%%%%%%%%%%%%%%%%%%%%%%%%%%%%%%%%%%
\subsection{Enthalpic allostery}
\label{subsec:enthalpy}
%%%%%%%%%%%%%%%%%%%%%%%%%%%%%%%%%%%%%%%%%%%%%%%%%%%%%%%%%%%%%%%%%%%%%%%%%%
%%%%%%%%%%%%%%%%%%%%%%%%%%%%%%%%%%%%%%%%%%%%%%%%%%%%%%%%%%%%%%%%%%%%%%%%%%

We use \eqref{SI:pu0um} and the expressions for $A$ and $B$
\eqref{SI:defA}, \eqref{SI:defB} to compute the interaction free energy
for enthalpic allostery.  The partition function of the free system
(DNA with no proteins bound) is
\begin{equation}
    Z_0 = \int \frac{du_0 du_m}{\pi} e^{-A(u_0^2+u_m^2)-2 B u_0 u_m} =
    \frac{1}{\sqrt{A^2-B^2}}
    \label{SI:Z0}
\end{equation}
where we divided by $\pi$ for convenience. 
The bulk free energy is then
\begin{eqnarray}
    F_0 = -k_B T \log Z_0 =
    \frac{k_BT}{2} \log \left( A^2 - B^2 \right)
\label{SI:F0}
\end{eqnarray}
Inserting a protein at site $u_0$ gives the partition function
\begin{eqnarray}
     Z_a &=& \int \frac{du_0 du_m}{\pi} \, e^{-A(u_0^2+u_m^2)-2 B u_0 u_m}
    e^{\beta h u_0} 
    \nonumber\\
    &=&
    \frac{1}{\sqrt{A^2-B^2}} \, \exp 
	\left[\frac{\beta^2 h^2 A}{4(A^2-B^2)}\right]
    \nonumber\\
    &=&
    Z_0 \, \exp \left[\frac{\beta^2 h^2 A}{4(A^2-B^2)}\right]
\end{eqnarray}
which, using $-k_B T \log Z_a = F_0 +\Delta F_a$ and \eqref{SI:F0}, 
gives the following excess free energy for a single bound protein
\begin{eqnarray}
    \Delta F_a = - \frac{\beta h^2}{4} \frac{A}{A^2-B^2}
    \label{SI:DFa_enth}
\end{eqnarray}
The partition function of two bound proteins is then
\begin{eqnarray}
    Z_{ab} &=& \int \frac{du_0 du_m}{\pi} \, e^{-A(u_0^2+u_m^2)-2 B u_0 u_m} 
    e^{\beta h (u_0 + u_m)} 
    \nonumber\\
    &=& \frac{1}{\sqrt{A^2-B^2}} \, 
	\exp \left[\frac{\beta^2 h^2}{2(A+B)}\right]
\end{eqnarray}
Using the relation
\begin{eqnarray}
    F_0 + \Delta F_a + \Delta F_b + \Delta\Delta F_{int} = -k_B T\log Z_{ab}
\end{eqnarray}
with \eqref{SI:F0} and \eqref{SI:Fa} (in our calculation the bound
proteins are assumed to be identical $\Delta F_a =\Delta F_b$) we get
the following estimate of the interaction free energy
\begin{eqnarray}
\boxed{
    \Delta\Delta F_{int}^E = \frac{\beta h^2}{2} \, \frac{B}{A^2-B^2}
    =  -h^2 S_m
}
\label{SI:DDFenth2}
\end{eqnarray}
where we have used Eqs.~\eqref{SI:defA} and \eqref{SI:defB} to obtain the
final result as a function of $S_m$.

%%%%%%%%%%%%%%%%%%%%%%%%%%%%%%%%%%%%%%%%%%%%%%%%%%%%%%%%%%%%%%%%%%%%%%%%%%
%%%%%%%%%%%%%%%%%%%%%%%%%%%%%%%%%%%%%%%%%%%%%%%%%%%%%%%%%%%%%%%%%%%%%%%%%%
\subsection{Entropic allostery}
\label{subsec:entropy}
%%%%%%%%%%%%%%%%%%%%%%%%%%%%%%%%%%%%%%%%%%%%%%%%%%%%%%%%%%%%%%%%%%%%%%%%%%
%%%%%%%%%%%%%%%%%%%%%%%%%%%%%%%%%%%%%%%%%%%%%%%%%%%%%%%%%%%%%%%%%%%%%%%%%%

We consider now the model for entropic allostery
\begin{equation}
    H = H_0 + \varepsilon \left( u_0^2 + u_m^2 \right)
\end{equation}
and compute the interaction free energy $\Delta\Delta F_{int}^S$ as a function
of the parameters $A$ and $B$ as done above for enthalpic allostery.
The partition function for a system with a single protein bound is
\begin{eqnarray}
    Z_a &=& \int \frac{du_0 du_m}{\pi} \, e^{-A(u_0^2+u_m^2)-2 B u_0 u_m} 
    e^{-\beta \varepsilon u_0^2} 
    \nonumber\\
    &=&  \frac{1}{\sqrt{A (A+\beta \varepsilon)-B^2}}
\end{eqnarray}
from which one can calculate the excess free energy due to the bound protein
\begin{equation}
    F_0 + \Delta F_a = -k_B T \log Z_a = 
    \frac{k_B T}{2} \log\left[A(A+\beta \varepsilon)-B^2 \right]
\end{equation}
Subtracting the expression for $F_0$ in \eqref{SI:F0} one gets 
the excess free energy associated to a single isolated protein
\begin{equation}
    \Delta F_a = \frac{k_BT}{2} \log \frac{A(A+\beta\varepsilon)-B^2}{A^2-B^2}
\label{SI:Fa}
\end{equation}
The partition function for two bound proteins is 
\begin{equation}
    Z_{ab} = \int \frac{du_0 du_m}{\pi} \, e^{-A(u_0^2+u_m^2)-2 B u_0 u_m} 
    e^{-\beta \varepsilon (u_0^2+u_m^2)}
\end{equation}
from which one gets the total free energy
\begin{equation}
    F_0 + 2 \Delta F_a + \Delta\Delta F_{int}^S = \frac{k_B T}{2} 
    \log\left[(A+\beta \varepsilon)^2-B^2 \right]
\label{SI:Fab}
\end{equation}
Subtracting from the previous expression \eqref{SI:F0} and \eqref{SI:Fa} one
obtains for the interaction free energy
\begin{eqnarray}
    \Delta\Delta F_{int}^S &=& \frac{k_BT}{2} 
    \log \frac{\left[\left(A+\beta\varepsilon\right)^2-B^2\right]
	\left(A^2-B^2\right)}
    {\left[A(A+\beta\varepsilon)-B^2\right]^2} 
    \nonumber \\
    &=& 
    \boxed{\frac{k_B T}{2} \log 
    \left[ 1 - \left( \frac{2\varepsilon S_m}
	{1+2\varepsilon S_0}\right)^2 \right]}
\label{SI:DDF}
\end{eqnarray}
where we have used the expressions for $A$ and $B$ \eqref{SI:defA},
\eqref{SI:defB}. There is no temperature dependence in the argument of the 
logarithm in \eqref{SI:DDF}, showing indeed that the interaction free energy 
is of purely entropic nature
\begin{equation}
    \Delta\Delta F_{int}^S = -T \Delta\Delta S_{int}
\end{equation}

%%%%%%%%%%%%%%%%%%%%%%%%%%%%%%%%%%%%%%%%%%%%%%%%%%%%%%%%%%%%%%%%%%%%%%%%%%
%%%%%%%%%%%%%%%%%%%%%%%%%%%%%%%%%%%%%%%%%%%%%%%%%%%%%%%%%%%%%%%%%%%%%%%%%%
\subsection{Mixed (enthalpic) allostery}
\label{subsec:mixed}
%%%%%%%%%%%%%%%%%%%%%%%%%%%%%%%%%%%%%%%%%%%%%%%%%%%%%%%%%%%%%%%%%%%%%%%%%%
%%%%%%%%%%%%%%%%%%%%%%%%%%%%%%%%%%%%%%%%%%%%%%%%%%%%%%%%%%%%%%%%%%%%%%%%%%

We repeat the analysis for the mixed case
\begin{eqnarray}
    H = H_0 + \varepsilon u_0^2 - h u_m
\end{eqnarray}
in which one protein induces a linear perturbation field, while the
other a quadratic one. This implies that the two proteins must be
different, as they interact with DNA in a substantial different way at
the two binding sites. The isolated proteins excess free energies were
already calculated previously and are given by Eqs.~\eqref{SI:DFa_enth}
and \eqref{SI:Fa}, respectively:
\begin{equation}
    \Delta F_a + \Delta F_b = - \frac{\beta h^2 A}{4(A^2-B^2)} +
    \frac{k_BT}{2} \log \frac{A(A+\beta\varepsilon)-B^2}{A^2-B^2}  
    \label{SI:isolated_mix}
\end{equation}
The partition function for two bound proteins is
\begin{eqnarray}
    Z_{ab} &=& \int \frac{du_0 du_m}{\pi} \, e^{-A(u_0^2+u_m^2)-2 B u_0 u_m} 
    e^{-\beta (\varepsilon u_0^2 + h u_m)}
    \nonumber \\
    &=& \frac{1}{\sqrt{A(A+\beta\varepsilon)-B^2}} \, 
    \exp 
    \left( 
    \frac{\beta^2 h^2}{4} \frac{A+\beta\varepsilon}{A(A+\beta\varepsilon)-B^2}
    \right)
    \nonumber\\
    \label{SI:zab-mixed}
\end{eqnarray}
From the above expression
\begin{eqnarray}
  -k_BT \log Z_{ab} = F_0 + \Delta F_a + \Delta F_b + \Delta \Delta F_{int}
\end{eqnarray}
and using \eqref{SI:F0}, \eqref{SI:zab-mixed} and 
\eqref{SI:isolated_mix} one gets 
\begin{eqnarray}
    \Delta\Delta F_{int}^M &=& \frac{\beta^2 h^2 \varepsilon}{4} \, 
    \frac{B^2}{\left(A^2-B^2 \right)\left[ A(A+\beta\varepsilon) - B^2 \right]}
    \nonumber \\
&=& \boxed{ \frac{\varepsilon h^2 S_m^2}{1+2\varepsilon S_0}}
    \label{SI:DDF_mixed}
\end{eqnarray}
where the expressions \eqref{SI:defA} and \eqref{SI:defB} were used.

%%%%%%%%%%%%%%%%%%%%%%%%%%%%%%%%%%%%%%%%%%%%%%%%%%%%%%%%%%%%%%%%%%%%%%%%%%%%
%%%%%%%%%%%%%%%%%%%%%%%%%%%%%%%%%%%%%%%%%%%%%%%%%%%%%%%%%%%%%%%%%%%%%%%%%%%%
\subsection{Global Allostery}
%%%%%%%%%%%%%%%%%%%%%%%%%%%%%%%%%%%%%%%%%%%%%%%%%%%%%%%%%%%%%%%%%%%%%%%%%%%%
%%%%%%%%%%%%%%%%%%%%%%%%%%%%%%%%%%%%%%%%%%%%%%%%%%%%%%%%%%%%%%%%%%%%%%%%%%%%

In the most general case a protein will exert both a linear and a
quadratic field. Here we consider the following model
\begin{eqnarray}
    H = H_0 + \varepsilon_a u_0^2 - h_a u_0 + \varepsilon_b u_m^2 - h_b u_m
\end{eqnarray}
using the same methodology as for the limiting cases above, we find the
following interaction free energy
\begin{eqnarray}
       \Delta\Delta F_{int}(m)&=&\frac{k_BT}{2}
        \log\left(1-\frac{4\varepsilon_a \varepsilon_b S_m^2}
	{(1+2\varepsilon_aS_0)   
        (1+2\varepsilon_bS_0)}\right) \nonumber\\
        &&-\frac{h_a h_b S_m}{1+2(\varepsilon_a+\varepsilon_b)S_0} \nonumber\\
        &&+\frac{\varepsilon_ah_b^2S_m^2}
	{(1+2(\varepsilon_a+\varepsilon_b)S_0)(1+2\varepsilon_bS_0)} \nonumber\\
       && +\frac{\varepsilon_b h_a^2 S_m^2}
	{(1+2(\varepsilon_a+\varepsilon_b)S_0)(1+2\varepsilon_aS_0)}
        \label{SI:general_allostery} 
\end{eqnarray}
The first term comprises the entropic contribution, whereas the latter
three are purely of an enthalpic nature. The second term couples the
two linear fields $h_a$ and $h_b$ similar to \eqref{SI:DDFenth2},
whereas the third and last term are the equivalent of mixed allostery
(see \eqref{SI:DDF_mixed}).

%%%%%%%%%%%%%%%%%%%%%%%%%%%%%%%%%%%%%%%%%%%%%%%%%%%%%%%%%%%%%%%%%%%%%%%%%%%%
%%%%%%%%%%%%%%%%%%%%%%%%%%%%%%%%%%%%%%%%%%%%%%%%%%%%%%%%%%%%%%%%%%%%%%%%%%%%
\subsection{Extended DNA-protein couplings}
\label{subsec:extended}
%%%%%%%%%%%%%%%%%%%%%%%%%%%%%%%%%%%%%%%%%%%%%%%%%%%%%%%%%%%%%%%%%%%%%%%%%%%%
%%%%%%%%%%%%%%%%%%%%%%%%%%%%%%%%%%%%%%%%%%%%%%%%%%%%%%%%%%%%%%%%%%%%%%%%%%%%

Thus far we have discussed the effect of proteins on a single local
site. In general a protein will interact with a few DNA sites and we
discuss here the case of extended DNA-protein couplings.

%%%%%%%%%%%%%%%%%%%%%%%%%%%%%%%%%%%%%%%%%%%%%%%%%%%%%%%%%%%%%%%%%%%%%%%%%%%%
\subsubsection{Enthalpic allostery: robustness of oscillating decay}
%%%%%%%%%%%%%%%%%%%%%%%%%%%%%%%%%%%%%%%%%%%%%%%%%%%%%%%%%%%%%%%%%%%%%%%%%%%%

We consider first the case of linear perturbations and consider the
following model
\begin{eqnarray}
    H = H_0 - \sum_{k=0}^{n_a-1} h_k u_k - 
   \sum_{l=n_a-1}^{n_a+n_b-2} h_{l+m} u_{l+m}
\label{SI:lin_extended}
\end{eqnarray}
with $n_{a,b}$ denoting the respective protein sizes. We allow the fields
$h_n$ to depend on the site. Discrete Fourier transform gives
\begin{equation}
    H = \frac{1}{2N} \sum_q \widetilde{K}_q |{\cal U}_q|^2 - 
    \frac{1}{N} \sum_q \left( C_q  + e^{2\pi i m q/N} D_q \right)  {\cal U}_q
\label{SI:extended}
\end{equation}
with
\begin{eqnarray}
 C_q &=& \sum_{k=0}^{n_a-1} h_k \, e^{2\pi i kq/N}   
 % \quad
 \label{SI:defC}
\\
 D_q &=& 
 \sum_{l=n_a-1}^{n_a+n_b-2} h_{l+m}  \, e^{2\pi i l q/N}
 \label{SI:defD}
\end{eqnarray}
Using the definition
\begin{eqnarray}
    h_q^* \equiv C_q + e^{2\pi im q/N} D_q
\end{eqnarray}
we can rewrite \eqref{SI:extended} as
\begin{eqnarray}
    H &=& 
    \frac{1}{2N} 
    \sum_q \widetilde{K}_q |{\cal U}_q|^2 
    - \frac{1}{2N} \sum_q \left(h_q^* {\cal U}_q + h_q {\cal U}_q^* \right)
    \nonumber\\
    &=&\frac{1}{2N} 
    \sum_q \widetilde{K}_q |{\cal U}_q'|^2 
    - \frac{1}{2N} \sum_q \frac{|h_q|^2}{\widetilde{K}_q}
    \label{SI:ext2}
\end{eqnarray}
where we have defined
\begin{eqnarray}
    {\cal U}_q' &=& {\cal U}_q - \frac{h_q}{\widetilde{K}_q}
\end{eqnarray}
To get the partition function we integrate on ${\cal U}_q'$.
The integration of the term $|{\cal U}_q'|^2$ gives $F_0$ the free
energy of the DNA.  The second term in \eqref{SI:ext2} contributes
to the excess free energies $\Delta F_a + \Delta F_b  + \Delta\Delta
F_{int}$. Expressing $|h_q|^2$ in terms of $C_q$ and $D_q$ we find
\begin{eqnarray}
    F_{ab} &=& F_0 - \frac{1}{2N} \sum_q 
	\frac{|C_q|^2 + |D_q|^2}{\widetilde{K}_q}
    \nonumber \\
    &&- \frac{1}{2N} \sum_q \frac{C_q D_{-q}+ C_{-q} D_q}
	{\widetilde{K}_q} \, e^{2\pi im q/N}
\end{eqnarray}
where we used the relations $C_q^*=C_{-q}$ and $D_q^*=D_{-q}$. The terms
proportional to $|C_q|^2$ and $|D_q|^2$ are $\Delta F_a$ and $\Delta
F_b$. The interaction term itself is given by:
\begin{eqnarray}
    \Delta\Delta F_{int} = -\frac{1}{2N} \sum_q 
    \frac{C_q D_{-q}+ C_{-q} D_q}{\widetilde{K}_q} \, e^{2\pi im q/N}
        \label{SI:ext3}
\end{eqnarray}
The localized perturbation limit, e.g. $-h(u_0+u_m)$, discussed 
in the main text is obtained by setting $n_a=n_b=1$ and $h_0=h_m=h$
in the two sums in \eqref{SI:defC} \eqref{SI:defD}, which 
corresponds to $C_q=D_q=h$ and hence
\begin{eqnarray}
    \Delta\Delta F_{int} = -\frac{h^2}{N} \sum_q 
    \frac{e^{2\pi im q/N}}{\widetilde{K}_q} 
\end{eqnarray}
which is Eq.~(11) of the main text.
We find therefore that an extended linear perturbation generates
a $q$-dependent term $C_q D_{-q} + C_{-q} D_q$ in the sum
\eqref{SI:ext3}. This does not modify the analytical structure
of the leading poles, unless $C_q D_{-q} + C_{-q} D_q$ vanishes
at $q=q_E$ as well, a very peculiar limiting case which we exclude
from our analysis. The conclusion is that the extended perturbation
\eqref{SI:lin_extended} generates the same asymptotic damped
oscillatory decay as the localized perturbation discussed in the text
\eqref{SI:Kqm-asympt}. This decay is thus robust.

%%%%%%%%%%%%%%%%%%%%%%%%%%%%%%%%%%%%%%%%%%%%%%%%%%%%%%%%%%%%%%%%%%%%%%%%%%%%
\subsubsection{Entropic and mixed allostery: lack of robustness of
oscillations}
%%%%%%%%%%%%%%%%%%%%%%%%%%%%%%%%%%%%%%%%%%%%%%%%%%%%%%%%%%%%%%%%%%%%%%%%%%%%

To test the effect of an extended coupling on entropic allostery 
we consider
\begin{eqnarray}
    H = H_0 + \varepsilon \left( u_0^2+u_1^2+u_{m+1}^2  
    \right)
\end{eqnarray}
We omit the details of the calculation which consists in computing the
probability distribution of three variables $p(u_0,u_1,u_{m+1})$ which
generalizes \eqref{SI:pu0um}. One then computes the interaction free
energy as done above for the other cases. The final result is
\begin{widetext}
\begin{eqnarray}
    \Delta\Delta F_{int} = \frac{k_BT}{2}  
    \log \left( 1- \frac{4\varepsilon^2 
	\left[1+2\varepsilon (S_0-S_1)\right](S_{m+1}^2 + S_m^2)
    +8\varepsilon^3 S_1 (S_{m+1} - S_m)^2}
    {[1+4\varepsilon^2 (S_0^2 - S_1^2) + 
	4 \varepsilon S_0](1+2\varepsilon S_0)} \right)
\label{SI:DDF-ent-extended}
\end{eqnarray}
\end{widetext}
We note that we can formally recover Eq.~\eqref{SI:DDF} 
(which we report here for convenience)
\begin{eqnarray}
    \Delta\Delta F_{int} = \frac{k_BT}{2} \log 
    \left[  1 - \left(\frac{2\varepsilon S_m}
	{1+2\varepsilon S_0}^2 \right) \right]
    \label{SI:very_last}
\end{eqnarray}
by setting $S_1=S_{\ec{m+1}}=0$ in \eqref{SI:DDF-ent-extended}. Also
\eqref{SI:DDF-ent-extended} describes a stabilizing interaction
$\Delta\Delta F_{int} \leq 0$. The main difference in the two cases
is that, while \eqref{SI:very_last} vanishes whenever $S_m=0$ and thus
has the same zeros as the enthalpic case, in \eqref{SI:DDF-ent-extended}
this does no longer happen. The $m$-dependence of the interaction comes
from terms proportional to $S_m^2$, $S_{m+1}^2$ and $S_m S_{m+1}$. 
The combination of these gives an overall exponential decay $\exp(-2m/\xi_E)$
with products of cosines of different phases which suppress the
oscillating part. The numerical solution of the general case
\begin{eqnarray}
    H = H_0 + \sum_{k=0}^{n_a-1} \varepsilon_k u_k^2 + 
    \sum_{l=n_a+1}^{n_a+n_b-2} \varepsilon_{m+l} u_{m+l}^2
\label{SI:quad_extended}
\end{eqnarray}
shows that oscillating components get more and more suppressed as the
number of interacting sites increases (larger $n_a$ and $n_b$), see
Fig.~2(b) in the main text.  Summarizing we expect for this case an
asymptotic behavior of the type
\begin{eqnarray}
    \Delta\Delta F_{int} \sim f(m) \, e^{-2m/\xi_E}
\end{eqnarray}
with $f(m)$ a very weak function of $m$. A very similar behavior holds
for the mixed case (see Fig.~2(c) in the main text).

%%%%%%%%%%%%%%%%%%%%%%%%%%%%%%%%%%%%%%%%%%%%%%%%%%%%%%%%%%%%%%%%%%%%  
\begin{figure}[t]
\includegraphics[width=0.42\textwidth,angle=0]{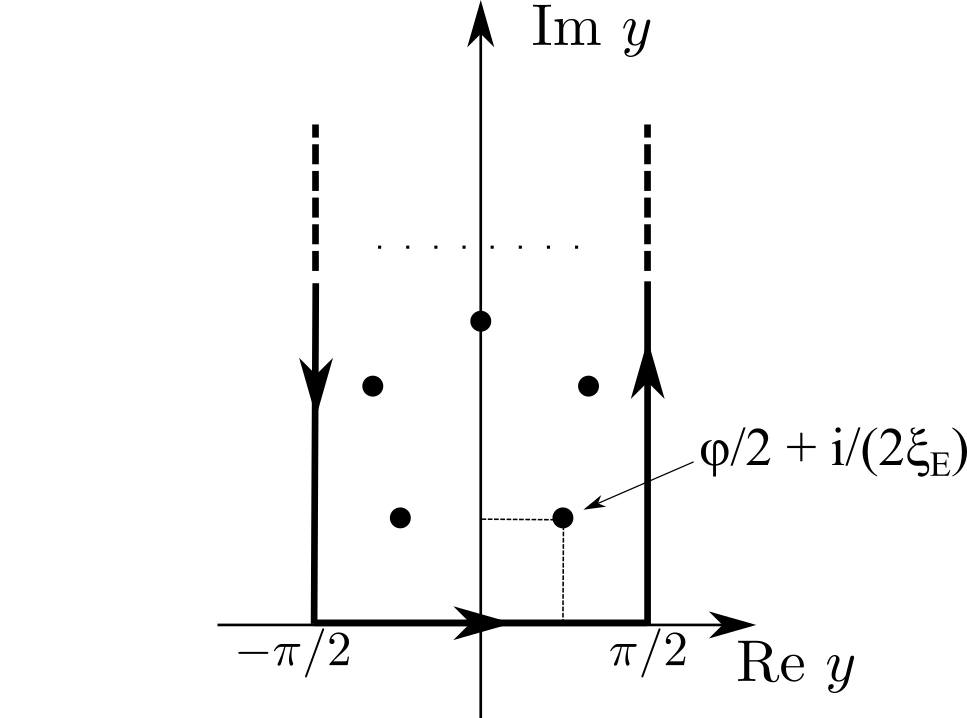}
\caption{Solid line: Integration contour for the calculation of the
integral \eqref{SI:enthalpy}. Circles: poles of the integrand obtained
by solving $\widetilde{K}_q=0$ (here $y\equiv \pi q/N$). We note that 
$\widetilde{K}_q$ is an even function of $q$. The 
leading large $m$ behavior is due to the pole(s) closest to the 
$Re(y)$ axis, is indicated with an arrow.}
\label{fig:contour}
\end{figure}
%%%%%%%%%%%%%%%%%%%%%%%%%%%%%%%%%%%%%%%%%%%%%%%%%%%%%%%%%%%%%%%%%%%%  

%%%%%%%%%%%%%%%%%%%%%%%%%%%%%%%%%%%%%%%%%%%%%%%%%%%%%%%%%%%%%%%%%%%%%%%%%%%
\section{Allosteric Length scale}
\label{sec:propagator}
%%%%%%%%%%%%%%%%%%%%%%%%%%%%%%%%%%%%%%%%%%%%%%%%%%%%%%%%%%%%%%%%%%%%%%%%%%%

%%%%%%%%%%%%%%%%%%%%%%%%%%%%%%%%%%%%%%%%%%%%%%%%%%%%%%%%%%%%%%%%%%%%%%%%%%%
%%%%%%%%%%%%%%%%%%%%%%%%%%%%%%%%%%%%%%%%%%%%%%%%%%%%%%%%%%%%%%%%%%%%%%%%%%%
\subsection{The propagator and its asymptotics}
%%%%%%%%%%%%%%%%%%%%%%%%%%%%%%%%%%%%%%%%%%%%%%%%%%%%%%%%%%%%%%%%%%%%%%%%%%%
%%%%%%%%%%%%%%%%%%%%%%%%%%%%%%%%%%%%%%%%%%%%%%%%%%%%%%%%%%%%%%%%%%%%%%%%%%%
In the limit $N \to \infty$ we can transform the discrete sum defining 
the propagator $S_m$ \eqref{SI:u0um-dir} into an integral
\begin{equation}
S_m = \frac{1}{N} \sum_q 
\frac{e^{2\pi i m q/N}}{\widetilde{K}_q}
= \frac{1}{\pi} \int_{-\pi/2}^{\pi/2} 
\frac{e^{2imy} \, dy}{\sum_n K_n \cos(2ny)}
\label{SI:enthalpy}
\end{equation}
where we used the change of variable $y \equiv \pi q/N$ and the expression
for $\widetilde{K}_q$ in (4). To evaluate the integral
one closes the integration domain in the complex y-plane, as shown in
Fig.~\ref{fig:contour}. Integrals along the two vertical lines cancel
each other by symmetry. The leading large-$m$ contribution comes from
the poles in the integrand with the smallest imaginary part (closest
to the $\text{Re}(y)$ axis). There cannot be poles on the real axis,
as this would lead to $\widetilde{K}_q=0$ for some real $q$, hence an
instability for that mode. Indicating the leading poles with

\begin{equation}
    y_\pm = \pm \frac{\phi}{2} + \frac{i}{2\xi_E}
\end{equation}
and defining $g(y) \equiv \sum_n K_n \cos(2ny)$ one gets
in the asymptotic limit ($m \gg 1$)
\begin{equation}
S_m \sim 2i \left[ \frac{e^{2imy_+}}{g'(y_+)}  + 
	\frac{e^{2imy_-}}{g'(y_-)} \right]
    \label{SI:Kqm-asympt}
\end{equation}
Defining $g'(y_+)=C_1+iC_2$ and using the relation 
$g'(y_+)=-[g'(y_-)]^*$ one has $g'(y_-)=-C_1+iC_2$ which gives 
Eq.(9) of the main text
\begin{equation}
    S_m \sim \Gamma \cos(m\phi + \phi_0) \, e^{-m/\xi_E}
\end{equation}
where 
\begin{equation}
    \Gamma = \frac{4}{\sqrt{C_1^2 + C_2^2}} \qquad
    \cos \phi_0 = \frac{C_2}{\sqrt{C_1^2 + C_2^2}}
\end{equation}

%%%%%%%%%%%%%%%%%%%%%%%%%%%%%%%%%%%%%%%%%%%%%%%%%%%%%%%%%%%%%%%%%%%%  
\begin{figure}[t]
\includegraphics[width=0.485\textwidth,angle=0]{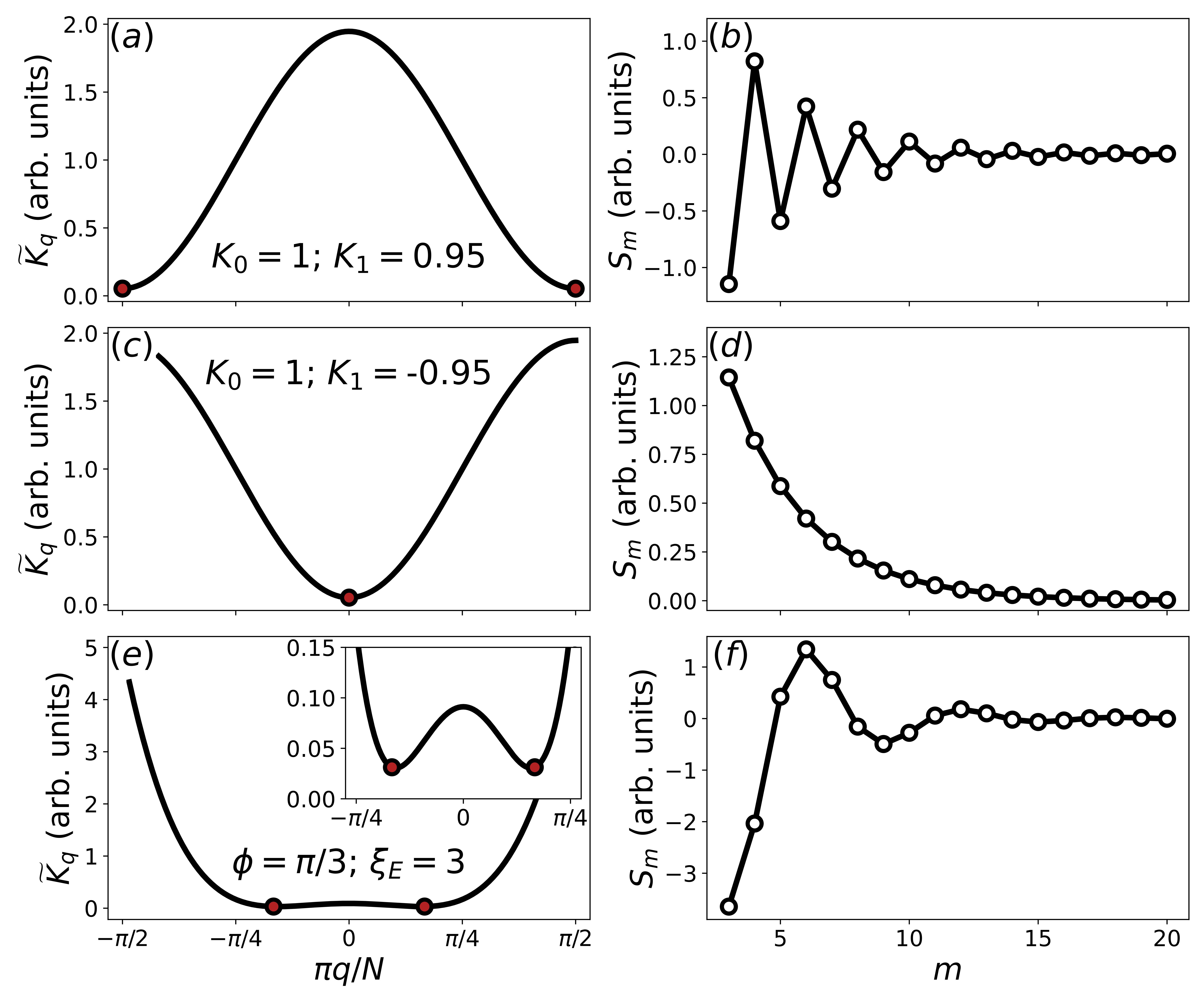}
\caption{Examples of $\widetilde{K}_q$ (left columns) with the
corresponding propagators $S_m$ vs. $m$ (right columns). The real parts 
of the poles of $\widetilde{K}_q$ are shown in (a,c,d) as red circles.
(a-d) correspond to a model with a single distal coupling $K_1$,
with $\widetilde{K}_q$ given in Eq.~\eqref{SI:Kq-minimal} and $K_1 >0$ 
(a-b), $K_1<0$ (c-d). In the case (a) the leading pole has a non-vanishing 
real component ($\phi=\pm \pi$) which produces oscillations with the 
highest possible frequency in the propagator $S_m \sim (-1)^m e^{-m/\xi}$, 
shown in (b). In the case (c) the leading pole has a vanishing real part
($\phi = 0$), which gives a monotonic decay $S_m \sim e^{-m/\xi}$, shown in (d). 
The case (e) corresponds to the continuum model \eqref{SI:minimal_model}. 
The real part of the pole $\mathrm{Re}(q_E)$ is close to the minimum 
of $\widetilde{K}_q$. Coarse-grained variables characterized by stiffnesses
as in (a) and (c) cannot be responsible for allosteric effects seen in DNA 
experiments, as they do not produce the damped oscillatory behavior.} 
\label{fig:examples}
\end{figure}
%%%%%%%%%%%%%%%%%%%%%%%%%%%%%%%%%%%%%%%%%%%%%%%%%%%%%%%%%%%%%%%%%%%%  

%%%%%%%%%%%%%%%%%%%%%%%%%%%%%%%%%%%%%%%%%%%%%%%%%%%%%%%%%%%%%%%%%%%%
\subsubsection{Some examples of $\widetilde{K}_q$}
%%%%%%%%%%%%%%%%%%%%%%%%%%%%%%%%%%%%%%%%%%%%%%%%%%%%%%%%%%%%%%%%%%%%

The values of $\phi$ and $\xi_E$ depend on the interaction parameters 
$K_n$. As a simple analytically solvable example we consider only one 
non-local term ($K_1 \neq 0$)
\begin{eqnarray}
\widetilde{K}_q = K_0 + K_1 \cos \left( \frac{2\pi q}{N} \right)  
\label{SI:Kq-minimal}
\end{eqnarray}
i.e. setting $K_l=0$ for $l \geq 2$. The stability condition
$\widetilde{K}_q > 0$ for real $q$ imposes $|K_1| < K_0$.  There
are thus two cases with $K_1$ positive or negative as shown in
Fig.~\ref{fig:examples}(a) and (\ec{c}). Defining $\omega \equiv \exp(2\pi
i q/N)$ one has
\begin{eqnarray}
    \cos \left( \frac{2\pi q}{N} \right) = \frac{1}{2} \left( \omega + 
    \frac{1}{\omega} \right)
\end{eqnarray}
and $\widetilde{K}_q = 0$ becomes a second degree equation for $\omega$
which give a single pole. For $0 < K_1 < K_0$ the pole is
\begin{eqnarray}
 \phi = \pm \pi, \quad \frac{1}{\xi_E} = 
 \log \left(  
 \frac{ K_0 + \sqrt{K_0^2-K_1^2}}{K_1}
 \right)
\label{SI:toy1}
\end{eqnarray}
while for $-K_0 < K_1 < 0$ the pole is
\begin{eqnarray}
 \phi = 0, \quad \frac{1}{\xi_E} = 
 \log \left(  
 \frac{ K_0 + \sqrt{K_0^2-K_1^2}}{|K_1|}
 \right)    
\label{SI:toy2}
\end{eqnarray}
We note that in both cases $\xi_E \to 0$ as $K_1 \to 0$, as allostery
requires some off-site couplings. Moreover the allosteric length diverges
as $|K_1| \to K_0$. In this limit the minimum $\min_q (\widetilde{K}_q)
\to 0$, the corresponding mode $q^*$ of the leading pole becoming
unstable.  In more complex situations, with further neighbor couplings
one needs to solve $\widetilde{K}_q = 0$ numerically, as this equation
is a higher degree polynomial in the variable $\omega$.

In the main text we introduced a minimal model for allostery with a
$\widetilde{K}_q$ given by 
\begin{equation}
    \widetilde{K}_q = C \left( q^2 - q_E^2\right)
	\left( q^2 - {q_E^*}^2\right) 
    = C\left( q ^4 - 2\mu q^2 + \lambda \right)
\label{SI:minimal_model}
\end{equation}
with $\mu=q_E^2+{q_E^*}^2 > 0$, $\lambda = |q_E|^4  > 0$ and
$C$ a constant. The quartic model \eqref{SI:minimal_model} is plotted in
Fig.~\ref{fig:examples}(e). This model is based on a $\widetilde{K}_q$
having a minimal number of zeros. Symmetry requires minimal
four such zeros, hence a polynomial of degree four.  We note that
\eqref{SI:minimal_model} is not a combination of cosines as in Eq.~(4)
of the main text. It describes instead a continuum model with energy:
\begin{equation}
    H_0 = C \int_0^L ds \left[ 
    \left( \partial^2_s u(s) \right)^2 - 2 \mu\left( \partial_s u(s) 
	\right)^2 +\lambda    u^2(s) \right]
\end{equation}
with $0 \leq s \leq L$ continuous coordinate (replacing the base pair
index) and $L$ the total length of the DNA.
The right column of Fig.~\ref{fig:examples} shows plots of the
propagators $S_m$ vs. $m$ corresponding to the stiffnesses shown in the
left column. As allosteric effects manifest themselves in experiments
as damped oscillations, the associated stiffness is expected to be
as in Fig.~\ref{fig:examples}(e).

%%%%%%%%%%%%%%%%%%%%%%%%%%%%%%%%%%%%%%%%%%%%%%%%%%%%%%%%%%%%%%%%%%%%
\subsection{Distal couplings in DNA models}
%%%%%%%%%%%%%%%%%%%%%%%%%%%%%%%%%%%%%%%%%%%%%%%%%%%%%%%%%%%%%%%%%%%%

The DNA-mediated allostery discussed in this paper is due to non-vanishing
distal couplings. Such couplings have been observed in simulations by
various groups \cite{lank00,noy12,lank09,esla11,skor21,voor23,fosa23}.
Their origin is twofold. Firstly, the coarsening of local degrees of
freedom leads to coarse-grained variables with couplings that extend
beyond nearest neighbors. Secondly, DNA interacts in a complex way with
solvent and counterions which leads to interactions which extend beyond
neighboring bases.

\begin{figure}[t]
\includegraphics[width=0.48\textwidth,angle=0]{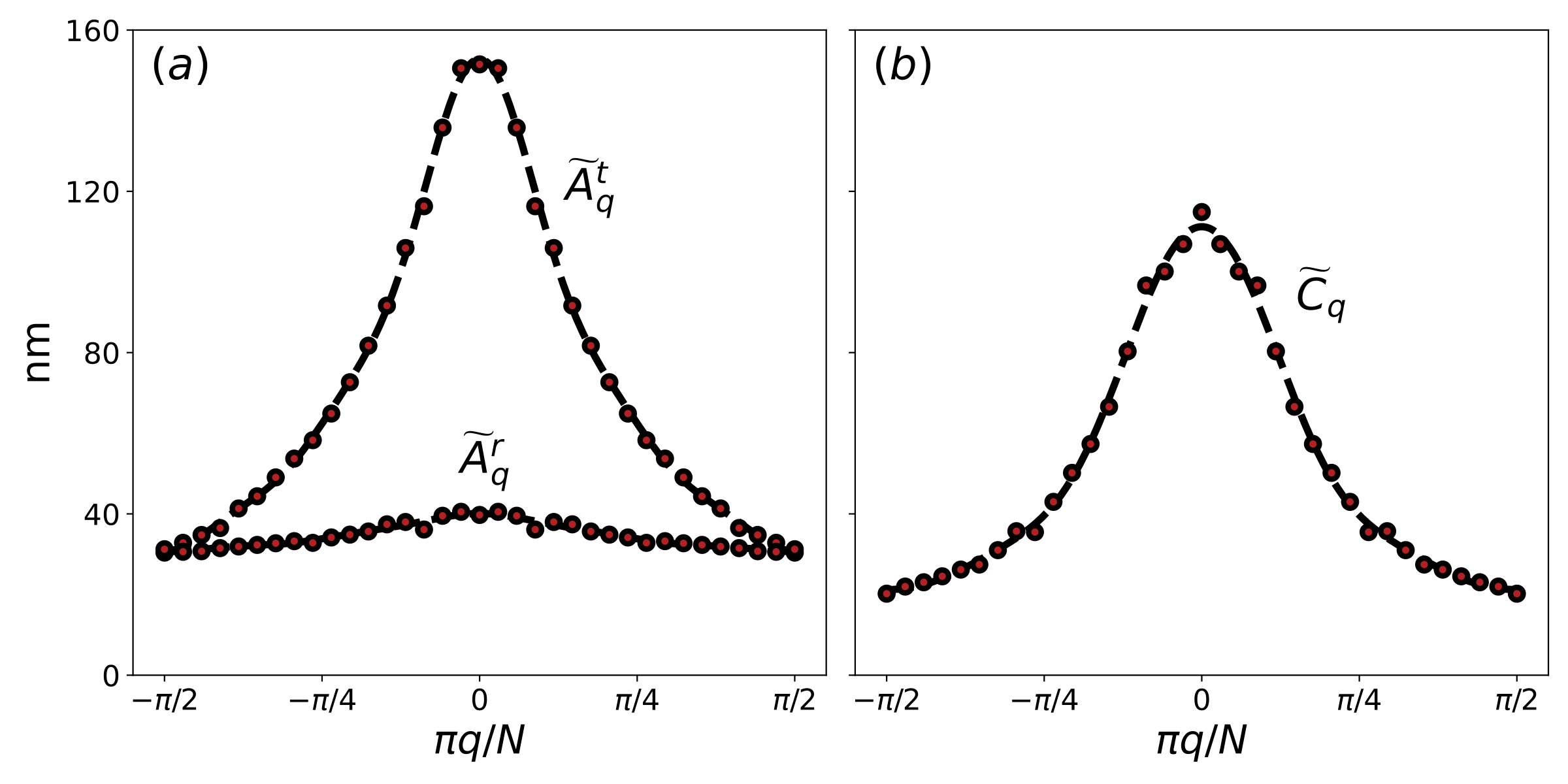}
\caption{Red circles: Fourier space stiffnesses for (a) bending
deformations (tilt $\widetilde{A}^t_q$ and roll $\widetilde{A}^r_q$)
and (b) twist deformation $\widetilde{C}_q$, as obtained from all
atom simulations averaging two different sequences of 44-mers of DNA.
Note that the very different behavior of the two bending stiffnesses
of Fig.~\ref{fig:stiff-twist-bendAA}(a): tilt has a very strong
$q$-dependence, while this is very weak in roll. This peculiar feature
was discussed in \cite{sege22}.  Dashed lines: Fourier decomposition of
the stiffnesses. The correlators $S_m$ associated to these degrees of
freedom alternate in sign as shown in Fig.~\ref{fig:examples}(b). Such a
behavior is not consistent with the oscillatory decay of DNA allostery
observed in experiment. Hence allosteric interactions are unlikely to
be carried by pure bending or twist deformations.}
\label{fig:stiff-twist-bendAA}
\end{figure}

The ordinary Twistable Wormlike Chain (TWLC) model, which describes
DNA deformations using twist and bending degrees of freedom, is a
very good model of DNA mechanics at sufficiently long length scales
($\sim100$~bp and beyond). Such model is strictly local and does not
have distal couplings, therefore it is not suitable to describe DNA at
a local scale where such couplings have been observed ($\sim 10$~bp).
Figure~\ref{fig:stiff-twist-bendAA} shows the Fourier space stiffness
of {\sl bending} and {\sl twist} modes as obtained from all-atom
simulations. The $q$-dependence indicates the presence of distal
couplings.  The stiffnesses have a global maximum at $q=0$ and minima
at $q= \pm N/2$ and are of the type shown in Fig.~\ref{fig:examples}(a).
As a consequence, the correlation functions of twist or bend deformations
decay with alternating signs, as shown in Fig.~\ref{fig:examples}(b). The
characteristic correlation lengths was estimated to be $\xi \approx 4$~bp
\cite{voor23}. Such degrees of freedom cannot produce an oscillatory
coupling free energy with the double helical period, as observed in
DNA allostery experiments \cite{kim13}. We can therefore conclude
that DNA allostery is unlikely to be carried by pure twist and bending
deformations.

%%%%%%%%%%%%%%%%%%%%%%%%%%%%%%%%%%%%%%%%%%%%%%%%%%%%%%%%%%%%%%%%%%%%  
%%%%%%%%%%%%%%%%%%%%%%%%%%%%%%%%%%%%%%%%%%%%%%%%%%%%%%%%%%%%%%%%%%%%  

In principle there are several other degrees of freedom describing DNA
deformations at the base pair level. A popular coarse grained model,
the rigid base model, introduces $12$ degrees of freedom per site
($6$ intra-basepair and $6$ inter-basepair, see e.g. \cite{petk14}).
In such model DNA bases A,T,C,G are treated as rigid bodies while
translation vectors and rotation angles are used to parametrize
their relative positions and orientations. The degrees of freedom
describing conformations of the rigid base model are also characterized
by non-local couplings and therefore have associated correlation
lengths (as shown by Eqs.~\eqref{SI:toy1} and \eqref{SI:toy2} a single
non-zero distal coupling is sufficient to induce a correlation length).
The oscillating interaction observed in experiments puts a constraint
on the coarse-grained variables acting as carriers of the allosteric
coupling. Such variables must necessarily have a stiffness of the
type of Fig.~\ref{fig:examples}(e). The analysis of simulation data
indicates that the major-groove width has the double-well shaped stiffness
which can potentially generate such oscillating decaying interactions.
Our model is quadratic in the variables $u_n$. Some proteins are known
to physically modify the DNA rather strongly so that anharmonic effects
may become relevant. There is currently little known about anharmonic
effects in DNA. These were recently explored for some degrees of freedom
\cite{voor23}, although the analysis is far from complete.  In that
paper a novel biasing scheme was introduced to investigate strongly
deformed DNA in all-atom simulations. The biasing is necessary as in
unconstrained simulations one does not observe anharmonic effects just
from thermal fluctuations.  Some degrees of freedom, such as twist, were
found to behave harmonically for a range to $\pm 10$ degrees per base pair
(see Fig.~2 of \cite{voor23}), which is a rather large value considering
it affects only two consecutive base pairs. Anharmonic effects could be
relevant in the enthalpic case as a quadratic perturbation $\sim u_m^2$
in our model suppresses fluctuations, thereby keeping the DNA deformations
within the harmonic regime. The enthalpic case with strongly DNA deforming
proteins will be analyzed in future work.

%%%%%%%%%%%%%%%%%%%%%%%%%%%%%%%%%%%%%%%%%%%%%%%%%%%%%%%%%%%%%%%%%%%%
\subsection{Allosteric lengths vs. persistence lengths}
%%%%%%%%%%%%%%%%%%%%%%%%%%%%%%%%%%%%%%%%%%%%%%%%%%%%%%%%%%%%%%%%%%%%

In principle any local degree of freedom with distal couplings has an
associated allosteric length $\xi_E$. The long length scale behavior 
of the DNA is described by torsional and bending persistence lengths.
The allosteric length is however unrelated to these persistence lengths
as it can be best illustrated using torsional elasticity as an example. 
The energy (in rescaled $k_BT$ units) is
\begin{eqnarray}
    \beta H_\mathrm{twist} &=& \frac{a}{2} \sum_{n=0}^{N-1} 
    \left[ C_0 \Omega_n^2 + \sum_{p=1}^L C_p (\Omega_n \Omega_{n+p} + 
    \Omega_n \Omega_{n-p})\right]
    \nonumber\\
    &=& \frac{a}{2N} \sum_q \widetilde{C}_q |\widetilde{\Omega}_q|^2,
    \label{ham}
\end{eqnarray}
where $\Omega_n$ is the excess twist per unit length between the two 
consecutive base pairs $n$ and $n+1$ and $a$ is a discretization length 
corresponding to the distance between base pairs ($a=0.34~nm$). $C_p$ 
indicate the stiffnesses and the \eqref{ham} is the same for as that
in the manuscript for the variable $u_n$. The torsional correlation 
function is given by (see details in \cite{skor21})
\begin{equation}
    \left\langle \cos \left( a \sum_{n=0}^{m-1} \Omega _n \right) \right\rangle 
    \stackrel{m \gg 1}{\sim} e^{-m a/\xi_T}, 
\end{equation}
with 
\begin{eqnarray}
        \xi_T &=&  2 \widetilde{C}_0 = 2\sum_{p=-L}^L C_p.
\end{eqnarray}
The calculation was done for correlated variables and it is a bit more complex than for the usual
Twistable Wormlike Chain (TWLC) where bending and twist at different sites are uncorrelated.
Hence in the model (\ref{ham}) the ordinary torsional correlation length is related to the $q=0$ 
(long wavelength limit) of the Fourier transform of the couplings. 
This can also be seen in the torque ensemble where the energy becomes:
\begin{eqnarray}
 \beta H &=& \beta H_\mathrm{twist} - \beta \tau a \sum_{n=0}^{N-1} \Omega_n
 \nonumber\\
 &=&  
 \frac{a \widetilde{C}_0}{2N} \widetilde{\Omega}_0^2 - \beta a \tau \widetilde{\Omega}_0 +
 \frac{a}{2N} \sum_{q \neq 0} \widetilde{C}_q |\widetilde{\Omega}_q|^2
\end{eqnarray}
which shows that the torque $\tau$ couples only to the mode $q=0$. The torque response 
is governed again by the $q=0$ component of the stiffness $\widetilde{C}_q$.

The allosteric effects we discussed in this paper are instead linked to the correlator
\begin{equation}
    \left\langle  \Omega _n \Omega_{n+m} \right\rangle \stackrel{m \gg 1}{\sim} 
    e^{-m a/\xi_E}
    \label{correl}
\end{equation}
where $\xi_E$ is determined from the leading (complex) solution of $\widetilde{C}_q=0$. 
$\xi_T$ and $\xi_E$ are therefore two distinct length scales, both linked 
to $\widetilde{C}_q$. 
From experiments probing DNA mechanical properties we know that $\xi_T \approx 200~nm 
= 600~bp$.
The other lengthscale is expected to be of the order of a few base pairs $\xi_E \approx 5~bp$
\cite{skor21}. The ordinary TWLC, considered the standard DNA model, describes 
DNA as uncorrelated twist and bending variables. It is a valid description at 
length scales $\gg \xi_E$. The previous discussion for simplicity was limited to twist, 
but for every local variable $u_n$ there is an associated correlation length $\xi_E$, 
defined from correlation functions as (\ref{correl}).

%%%%%%%%%%%%%%%%%%%%%%%%%%%%%%%%%%%%%%%%%%%%%%%%%%%%%%%%%%%%%%%%%%%%  
\begin{figure}[t]
\includegraphics[width=0.40\textwidth,angle=0]{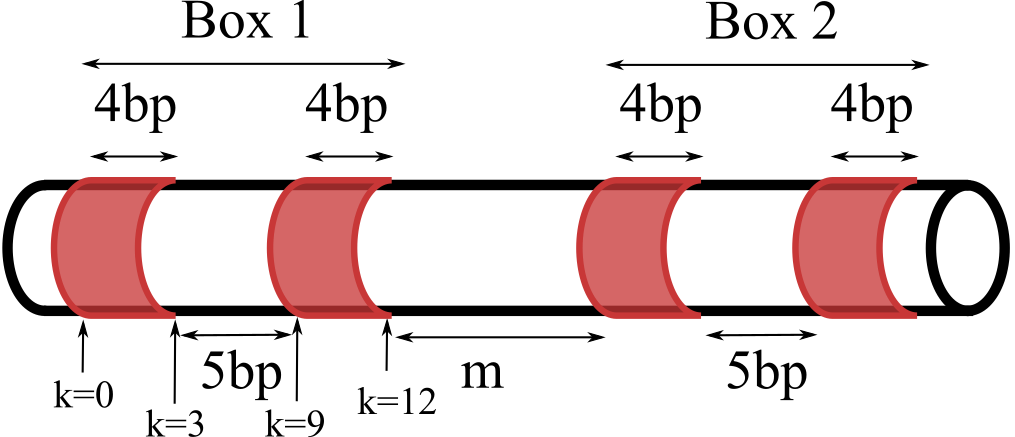}
\caption{Schematic view of the structure of the ComK-DNA binding interface 
(red dashed area). There are two ``binding boxes'' separated by a linker of 
$m$ base-pairs. Each box consists of two regions of fixed sequence of four 
base pairs each, separated by a $5$~bp variable region. Two proteins bind 
at each binding box, so allostery involves in total four ComK. We model 
the proteins binding in each box using linear and quadratic fields in each 
of the $8$ binding sites, as given in Eq.~\eqref{SI:int-ComK}. The $4$~bp 
DNA sequences are mirror symmetric, hence we use the same field values at 
sites $k=0,12$, $k=1,11$, $k=2,10$ and $k=3,9$.}
\label{fig:ComK-bind}
\end{figure}
%%%%%%%%%%%%%%%%%%%%%%%%%%%%%%%%%%%%%%%%%%%%%%%%%%%%%%%%%%%%%%%%%%%%  

%%%%%%%%%%%%%%%%%%%%%%%%%%%%%%%%%%%%%%%%%%%%%%%%%%%%%%%%%%%%%%%%%%%%%%%%%%%%%%
%%%%%%%%%%%%%%%%%%%%%%%%%%%%%%%%%%%%%%%%%%%%%%%%%%%%%%%%%%%%%%%%%%%%%%%%%%%%%%
\section{Fitting procedure of ComK data}
\label{supp:fitComK}
%%%%%%%%%%%%%%%%%%%%%%%%%%%%%%%%%%%%%%%%%%%%%%%%%%%%%%%%%%%%%%%%%%%%%%%%%%%%%%
%%%%%%%%%%%%%%%%%%%%%%%%%%%%%%%%%%%%%%%%%%%%%%%%%%%%%%%%%%%%%%%%%%%%%%%%%%%%%%
We discuss here the procedure used to fit the ComK data of Fig.~3(b) 
of the main text. The comK binding sites
consists of two boxes as shown in Fig.~\ref{fig:ComK-bind}. Each box has
two $4$~bp regions of conserved sequences throughout various organisms,
which are separated by a variable $5$~bp tract \cite{rose21}. Two
ComK molecules bind to each binding box, so that in total four ComK
are involved in the allosteric communication \cite{rose21}.  To fit
the experimental data, we model the binding of a first pair of ComK
molecules at a binding box by two extended perturbations each spanning
over four sites in order to assimilate the length of the binding domains
within one box. The perturbations are placed at a fixed distance
of five sites from one another accounting for the five base-pairs
between binding domains. Each of the individual perturbations entails
a linear field $(h_0,h_1,h_2,h_3)$ and a quadratic field $(\epsilon_0,
\epsilon_1,\epsilon_2,\epsilon_3)$ such that the free energy of DNA
bound to a single pair of ComK is given by:
\begin{equation}
H = H_0 + \sum_{k=0}^3 \left[ \varepsilon_k \left(u_k^2 + u_{12-k}^2 \right) 
- h_k \left(u_k + u_{12-k} \right) 
\right]
\label{SI:int-ComK}
\end{equation}

Where the $12-k$ accounts for the anti-symmetric binding of the two ComK
molecules.  The binding of a second pair of ComK proteins is modelled
by the same perturbation displaced over a distance of $m$ sites. For a
given combination of $h_k$ and $\epsilon_k$, $\Delta\Delta F_{int}(m)$
is calculated numerically. In order to find an appropriate choice of
$h_k$ and $\epsilon_k$ fitting the data we implemented a Monte Carlo
algorithm. During each MC step $h_k$ and $\epsilon_k$ are altered
after which the change is accepted/rejected based on the Metropolis
acceptance criterion.  Due to the presence of linear and quadratic fields
the interaction free energy is comprised of an energetic (Enthalpic and
Mixed allostery) and an entropic component.  We allow for the entropic
term to be multiplied by an integer prefactor reflecting the relevance
of several degrees of freedom in DNA-ComK binding.

%%%%%%%%%%%%%%%%%%%%%%%%%%%%%%%%%%%%%%%%%%%%%%%%%%%%%%%%%%%%%%%%%%%%  
\begin{figure}[t]
\includegraphics[width=0.42\textwidth,angle=0]{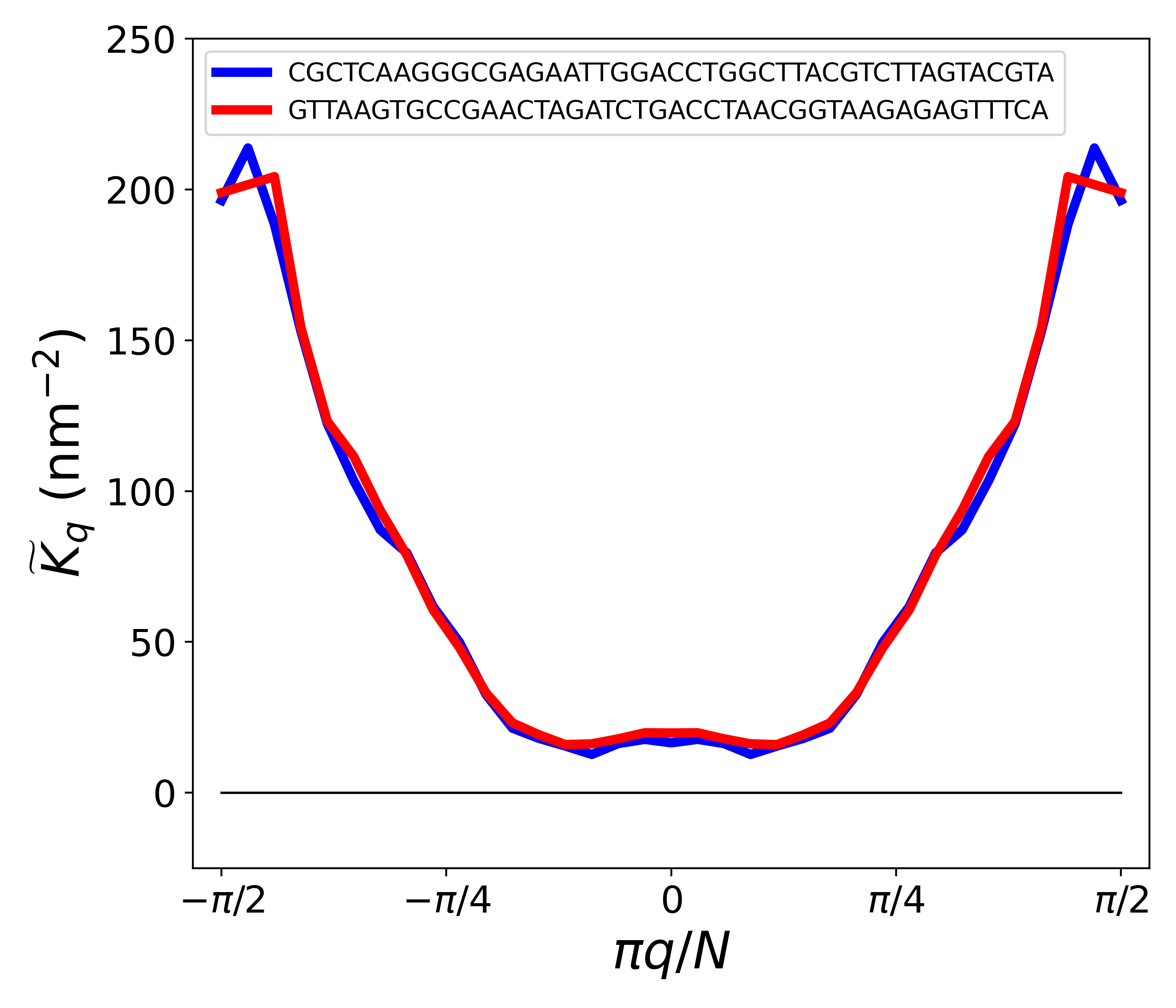}
\caption{Plot of the momentum-space stiffness for the major groove 
width obtained from all-atom simulations for two oligomers of 44 bps. 
The two sequences are given in the top of the figure.}
\label{fig:Kq_seq}
\end{figure}
%%%%%%%%%%%%%%%%%%%%%%%%%%%%%%%%%%%%%%%%%%%%%%%%%%%%%%%%%%%%%%%%%%%%  

%%%%%%%%%%%%%%%%%%%%%%%%%%%%%%%%%%%%%%%%%%%%%%%%%%%%%%%%%%%%%%%%%%%%%%%%%%%%%%
%%%%%%%%%%%%%%%%%%%%%%%%%%%%%%%%%%%%%%%%%%%%%%%%%%%%%%%%%%%%%%%%%%%%%%%%%%%%%%
\section{All atom simulations}
\label{supp:AA}
%%%%%%%%%%%%%%%%%%%%%%%%%%%%%%%%%%%%%%%%%%%%%%%%%%%%%%%%%%%%%%%%%%%%%%%%%%%%%%
%%%%%%%%%%%%%%%%%%%%%%%%%%%%%%%%%%%%%%%%%%%%%%%%%%%%%%%%%%%%%%%%%%%%%%%%%%%%%%

All simulations were done using version 2020.4 of Gromacs \cite{gromacs}
and the Amberff99 parmbsc1 force field (based on the parm99 forcefield,
improved with bsc0 \cite{pere07} corrections and bsc1 corrections
\cite{ivan16}).  Water was modelled using the TIP-3P model \cite{jorg83}
non-bonded interactions were cutoff at 1.0 nm and PME Mesh Ewald
interactions was used for electrostatics.  Simulations started from two
44bp random sequences. The starting configuration was generated using
3DNA \cite{li19}. The DNA molecules were placed into a dodecahedral box,
leaving 2.0 nm on either side of the molecule, with periodic boundary
conditions and solvated in a 150 mM NaCl solution after which the overall
charge in the system was neutralised. This structure was energy minimised
with a tolerance of 1000 kJ/mol to make sure no overlap remained between
solvent molecules and DNA.  Subsequently, the molecule was equilibrated in
the NVT ensemble for 100 ps where temperature was kept at $300$~K using
a velocity rescaling thermostat \cite{buss07} and then equilibrated
for another $100$~ps in the NPT ensemble at the same temperature but
using a Parrinello-Rahman barostat \cite{parr81} to fix the pressure at
$1.0$~bar. Production runs were performed under the same conditions for
100 ns for each sequence. All simulations were performed using a $2$~fs
time step in a leapfrog integrator, using LINCS \cite{hess97} to constrain
the covalent bonds involving hydrogen atoms. Afterwards the trajectories
were analyzed and the groove widths extracted using Curves+ \cite{lave09}. 

% \ms{The resulting momentum-space stiffness for the two simulated oligonucleotides 
% are shown in Fig. \ref{fig:Kq_seq}.}

%%%%%%%%%%%%%%%%%%%%%%%%%%%%%%%%%%%%%%%%%%%%%%%%%%%%%%%%%%%%%%%%%%%%%%%%%%%%%%%%%%%%%%%%%%%%%%%%%%%%%
%%%%%%%%%%%%%%%%%%%%%%%%%%%%%%%%%%%%%%%%%%%%%%%%%%%%%%%%%%%%%%%%%%%%%%%%%%%%%%%%%%%%%%%%%%%%%%%%%%%%%
\begin{table}[t]
\renewcommand{\arraystretch}{1.5}
\centering
    \begin{tabular*}{\linewidth}{l@{\extracolsep{\fill}}|c|ccccc}
        \hline
        \hline
        \hspace{0.1cm} & $K_0$ & $K_1$ & $K_2$ & $K_3$ & $K_4$ & $K_5$\\
        \hline
        \hspace{0.2cm}$\widetilde{K}_q$ (major-groove width)   & 77 & -43 & 15 & -3.5 & 3.8 & -1.2 \\
        \hline
        \hspace{0.2cm}$\widetilde{A}^t_q$ (tilt)  & 74 & 26 & 7.3 & 4.0 & 1.5 & 0.9 \\
        \hspace{0.2cm}$\widetilde{A}^r_q$ (roll)  & 35 & 2 & 0.4 & 0.2 & 0.01 & 0.1 \\
        \hline
        \hspace{0.2cm}$\widetilde{C}_q$   (twist) & 53 & 20 & 6.6 & 2.3 & 0.2 & -0.1 \\
        \hline
        \hline
    \end{tabular*}
    \caption{Real space stiffnesses $K_m$ for the major-groove width
    $\widetilde{K}_q$, tilt $\widetilde{A}^t_q$, roll $\widetilde{A}^r_q$
    and twist $\widetilde{C}_q$.  Values are obtained by fitting
    the q-stiffnesses extracted from all-atom data (see Figures~4
    and \ref{fig:stiff-twist-bendAA}) with Eq.~3.  In contrast to
    bending and twist deformations, the off-site couplings $K_p$ for
    the major-groove width have alternating signs, giving rise to the
    distinctive double-welled shape of $\widetilde{K}_q$.}
    \label{tab:real_space_couplings}
\end{table}
%%%%%%%%%%%%%%%%%%%%%%%%%%%%%%%%%%%%%%%%%%%%%%%%%%%%%%%%%%%%%%%%%%%%%%%%%%%%%%%%%%%%%%%%%%%%%%%%%%%%%
%%%%%%%%%%%%%%%%%%%%%%%%%%%%%%%%%%%%%%%%%%%%%%%%%%%%%%%%%%%%%%%%%%%%%%%%%%%%%%%%%%%%%%%%%%%%%%%%%%%%%
Figure~\ref{fig:Kq_seq} shows the $\widetilde{K}_q$ as calculated
from the two individual $44$-mers random sequences. The stiffness is
weakly dependent on the sequence, suggesting the groove width deformations
can be approximately described by a homogeneous model. The sequence to 
sequence variation for bending and twist deformation is typically much 
stronger, see simulation data in \cite{skor21}.
We report in Table~\ref{tab:real_space_couplings} the local stiffness
($K_0$) and the distal couplings ($K_1$, $K_2$, $K_3$ \ldots) as obtained from
all atom simulations data analysis for major groove width, tilt, roll and 
twist coordinates. 

To produce a propagator $S_m$ with a monotonic decay or a decay with
alternating signs (as those shown in Fig.~\ref{fig:examples} (b)
and (d)) a single distal coupling $K_1 \neq 0$ is sufficient. One
can think about the equivalent problem in the one-dimensional
ferromagnetic or antiferromagnetic Ising model with nearest-neighbor
coupling. Equations~\eqref{SI:toy1} and ~\eqref{SI:toy2} give the
analytical expressions of $\xi_E$ in these two cases.  To produce damped
oscillations with the period of the double helix longer range couplings
are necessary. It is likely that all-atom simulations do not reproduce
these types of effects because of either limitations in the reachable
timescales ($\sim \mu s$) or because of problems with the parametrization.
Current all-atom DNA models are parametrized using classical force fields
which have some shorcomings, as discussed recently in \cite{lieb23}.
One of the conclusions of that paper is that current state-of-the-art
force fields overstabilize nucleic acids suppressing possible ``higher
degree conformational transitions''.  These conformational changes
may lead to stronger correlation effects and thus to closer agreement
with allostery experiments. There is an ongoing effort to develop more
accurate force fields for nucleic acids \cite{lieb23} and allostery data
could be used as test case.

\bibliography{references}
\end{document}